\documentclass[pra,aps,letterpaper,longbibliography,twocolumn,12point]{revtex4-1}

\usepackage{graphicx}
\usepackage{dcolumn}
\usepackage{bm}
\usepackage{color}      

\usepackage{amstext}
\usepackage{amsmath}
\usepackage{graphics}
\usepackage{amssymb}
\usepackage{color}

\newcommand{\bra}[1]{\langle #1|}
\newcommand{\ket}[1]{|#1\rangle}

\newcommand \be{\begin{equation}}
\newcommand \ee{\end{equation}}
\newcommand \bea{\begin{eqnarray}}
\newcommand \eea{\end{eqnarray}}
\newcommand \nn{\nonumber}

\newcommand \bfn{{\bf n}}
\newcommand \bfomega{{\boldsymbol \omega}}

\begin{document}

\title{Rydberg mediated entanglement in a two-dimensional neutral atom qubit array} 

\author{T. M. Graham, M. Kwon, B. Grinkemeyer, Z. Marra, X. Jiang}
\author{M. T. Lichtman}
\altaffiliation{Present address: 
Joint Quantum Institute and Department of Physics,
University of Maryland, College Park, MD 20742}
\author{Y. Sun}
\altaffiliation{Present address: Interdisciplinary Center for Quantum Information,
National University of Defense Technology, Changsha 410073, China}
\author{M. Ebert}
\altaffiliation{ColdQuanta, Inc., 811 E. Washington Ave, Suite 408, Madison, WI 53703}
\author{M. Saffman}
\altaffiliation{ColdQuanta, Inc., 811 E. Washington Ave, Suite 408, Madison, WI 53703}
\affiliation{
Department of Physics, University of Wisconsin-Madison, 1150 University Avenue, Madison, Wisconsin 53706
}

\begin{abstract}
We demonstrate high fidelity two-qubit Rydberg blockade and entanglement in a two-dimensional qubit array. The qubit array is defined by a grid of blue detuned lines of light with 121 sites for trapping  atomic qubits. Improved  experimental methods have increased the observed Bell state fidelity to $F_{\rm Bell}=0.86(2)$. Accounting for errors in state preparation and measurement (SPAM) we infer a fidelity of $F_{\rm Bell}^{\rm -SPAM}=0.88$. Accounting for errors in single qubit operations we infer that 
 a Bell state created with the Rydberg mediated $C_Z$ gate has a fidelity of $F_{\rm Bell}^{C_Z}=0.89$. Comparison with a detailed error model based on quantum process matrices indicates that  finite atom temperature  and laser noise are the dominant error sources contributing  to the observed gate infidelity.
\end{abstract}

\date{\today}

\maketitle

Achieving the promise of a computational advantage for quantum machines is predicated on the development of approaches that combine a large number of qubits with a high fidelity universal gate set. A broad range of experimental platforms for quantum computing are being developed\cite{Ladd2010} and very high fidelity  two-qubit gates have been implemented in trapped ion and superconducting systems with small numbers of qubits: $F_{\rm Bell}\ge 0.999$ with two trapped ions\cite{Ballance2016,*Gaebler2016} and
a phase gate fidelity $F_{C_Z} > 0.99$ with five superconducting qubits\cite{Barends2014}.  As the number of qubits in a quantum computer is scaled up, crosstalk and undesired  interactions may limit fidelity. Average Bell state fidelities of $F_{\rm Bell}= 0.975$ were obtained in an 11 qubit ion trap\cite{Wright2019}. An approach based on  encoding in hyperfine states of optically trapped neutral atoms holds great promise for scaling the number of qubits without limiting gate fidelities. The physical attribute that enables scaling with low crosstalk is the separation by 12 orders of magnitude between the weak coupling strength of neutral atom hyperfine qubits, and the strong interactions of Rydberg excited atoms\cite{Saffman2010} that are used to realize entangling gates\cite{Jaksch2000}. We report here on experimental progress 
in achieving high entanglement fidelity in a 2D array.  Since these results were measured in a large array of qubits (121 traps with an average of 55\% loading), the architecture is compatible with scaling to algorithms involving many  qubits. The Bell state fidelity of $F_{\rm Bell}^{C_Z}= 0.89$, together with previously demonstrated single qubit gates with $F>0.99$\cite{Xia2015,YWang2016}, and atom rearrangement capabilities\cite{Endres2016,*Barredo2016,*Kumar2018,*Barredo2018} suggest that neutral atom arrays will soon be capable of advancing the state of the art in gate based quantum computing. 

A computationally universal set of quantum gates can be built from one- and two-qubit operations. High fidelity one-qubit gates with fidelities determined by randomized benchmarking exceeding 0.99 and crosstalk to other sites less than 0.01  have been demonstrated in 2D\cite{Xia2015} and 3D\cite{YWang2016} arrays of neutral atom qubits. However the fidelity of two-qubit entangling gates has been limited to much lower values. The highest  fidelity results from the last few years for entanglement of pairs of neutral atoms are 0.79\cite{Maller2015}, 0.81\cite{Jau2016}, 0.59\cite{YZeng2017},  0.81\cite{Picken2019}. These fidelity numbers are corrected for SPAM errors and in some cases also atom loss. Recent progress with qubits encoded in one hyperfine ground state and one Rydberg state has demonstrated entanglement fidelity of 0.97\cite{Levine2018}, although the use of Rydberg encoding limits the coherence time to $<0.1 ~\rm ms$, which is much shorter than the seconds of coherence time that have been achieved with qubits encoded in hyperfine ground states\cite{YWang2016,CSheng2018}. 

\begin{figure}[!t] 
  \includegraphics[width=1.\columnwidth]{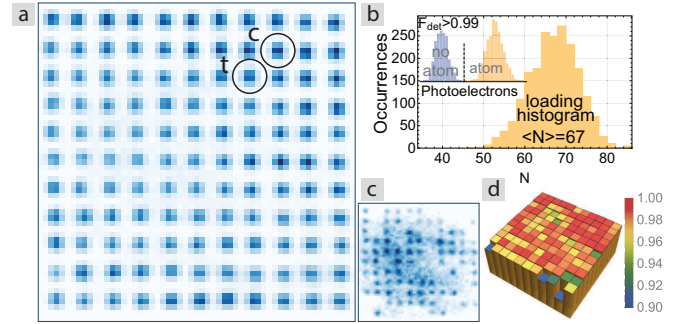}
  \vspace{-.6cm}
  \caption{(color online) Atomic qubit array. a) Averaged fluorescence image of 121 site array after ICA processing with the control and target sites used for the presented data labeled. b) Loading histogram showing an average filling fraction of 55\% with inset showing state detection inferred after blow away of atoms in $f=4$. Accounting for all 121 sites the state detection error found from the overlap of Gaussians fitted to the $\ket{0}$ and $\ket{1}$ distributions was mean$=0.014$, median$=0.003$. c) Single instance of atom loading. d) Atom retention probability after measurement.  The atom retention probability after two subsequent readout sequences was mean$=96.9\%$, median$=97.9\%$.  The difference between mean and median is due to some sites on the edge of the array having lower retention due to optical aberrations.   }
  \label{fig.array}
  \vspace{-.5cm}
\end{figure}

The experimental setup is an upgraded version of that described in \cite{Maller2015}. A two-dimensional array of Cs atoms is prepared using a projected optical lattice with period $d=3.1(1)~\mu\rm m$ (numbers in parentheses are uncertainties in the last digit) and wavelength $\lambda=825~\rm nm$.  (see Fig.\ref{fig.array}). In contrast to our previous work with a Gaussian beam array\cite{Piotrowicz2013}, the array structure is defined by a square grid of lines of light that are prepared using diffractive optical elements\cite{Lichtman2015}. Each unit cell  provides 3D atom confinement with the transverse localization due to the repulsive walls of blue-detuned light, and axial confinement perpendicular to the plane of the array provided by diffractive spreading of the lines. We measure vacuum limited lifetimes of $\sim 30~\rm s$, longitudinal coherence times in the Cs clock states of  $T_1=0.75~\rm s$ and an average temperature of $T_a\simeq 15~\mu\rm K$. The atomic temperature implies a limit on the clock state coherence time due to motional variation of the trap light intensity of $T_2^*=1.6~\rm ms$.  

The array is prepared by combining four laser sources with different frequencies such that the four beams defining each unit cell are separated by many MHz, but the frequencies are repeated in neighboring cells. With this configuration the structure and position of each trapping site are insensitive to phases caused by variations in optical path length which provides a very stable array. However,  Talbot interference still occurs leading to additional trapping planes at axial separations of  $L=2(2d)^2/\lambda=93~\mu\rm m$.   Detection of atoms in the array  is hampered by a diffuse background of scattering  from atoms in the additional Talbot planes. We effectively reduce the background noise with regions of interest for each trap site, that are determined using an independent component analysis (ICA) algorithm\cite{Lichtman2015}, see Fig. \ref{fig.array}.
Alternatively the Talbot planes can be eliminated by making each line a different frequency. We have implemented this using acousto-optic deflectors to create the lines and thereby generated arrays with up to 196 trapping sites and an average of 110 trapped atoms. Details of this approach will be given elsewhere\cite{Graham2019b}.

\begin{figure}[!t] 
  \includegraphics[width=.85\columnwidth]{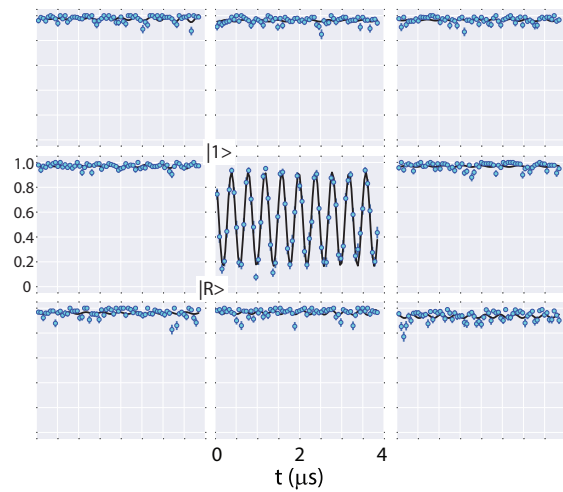}
\vspace{-.3cm}
  \caption{(color online) Single site ground-Rydberg Rabi oscillations at $\Omega_R/2\pi=2.5~\rm MHz$ with negligible crosstalk to the surrounding eight sites (see \cite{SM2019} for analysis).  All atoms in the array were prepared in $\ket{1}$ directly before the targeted site was illuminated with Rydberg beams.  }
  \label{fig.RabiGR}
   \vspace{-.5cm}
\end{figure}

The trapped Cs atoms are optically pumped into the clock states which form a qubit basis of $\ket{0}=\ket{6s_{1/2},f=3,m_f=0}$
and $\ket{1}=\ket{6s_{1/2},f=4,m_f=0}.$ State $\ket{1}$ is resonantly coupled to the Rydberg state $\ket{R}=\ket{66s_{1/2},m_j=-1/2}$  using a two-photon transition with counterpropagating $\lambda_1=459$ nm $(\sigma_+)$ and $\lambda_2=1038$ nm $(\sigma_-)$ laser beams which couple $6s_{1/2}\rightarrow 7p_{1/2}\rightarrow 66s_{1/2}.$ We detune by $+680$  MHz from the center of mass of the $7p_{1/2}$ state and use a magnetic bias field of $0.6~\rm mT$ directed along the quantization axis, which is collinear with the beam $\bf k$ vectors, to separate the Rydberg $m_j=\pm 1/2$ states. The choice of Rydberg principal quantum number is lower than in our previous demonstrations and is desirable for minimizing perturbations from background electric fields. With the small array period used here there is sufficient blockade strength and state lifetime at $n=66$ that the resulting errors are minor contributions to the overall error model (see  Table \ref{tab.budget} below).

The Rydberg excitation beams are focused to beam waists ($1/e^2$ intensity radii) of $w_{1}=w_2=3.0~\mu\rm m$ which 
 are pointed to address desired sites in the array using two crossed acousto-optic modulators for each beam.  By using acousto-optic modulators to point Rydberg lasers, it is possible to address different sites in the array with sub-microsecond switching times which will make possible running multi-qubit algorithms.   Figure \ref{fig.RabiGR} shows ground to Rydberg Rabi oscillation data in the array.
The trap light is turned off during the Rydberg pulse. Since the blue detuned array traps Rydberg atoms when turned on again\cite{SZhang2011,Zhang2012thesis} there is only minimal mechanical loss of Rydberg states. In order to detect Rydberg excitation we turn on a short 9.2 GHz microwave pulse (duration 70 $\mu\rm s$) to photoionize the Rydberg atom. Rydberg detection efficiencies are typically 80-90\%. Rydberg excitation was performed using diode lasers that are stabilized to high finesse optical resonators ($\sim  5 ~\rm kHz$ linewidth)\cite{SM2019}. It has been recognized that phase noise of diode lasers contributes to decay of ground-Rydberg oscillations\cite{deLeseleuc2018}, with improved performance achieved by resonator filtering\cite{Levine2018}. Here we demonstrate comparable performance, without any resonator filtering, but with careful tuning of the electronic Pound-Drever-Hall lock parameters to reduce the amplitude of servo bumps.

A  curve fit to the Rabi oscillations at the selected site in Fig. \ref{fig.RabiGR} does not reveal a statistically significant  decay time. The radiative lifetime of the $66s$ state is  $130 ~\mu\rm s$ and the motional dwell time of a Rydberg atom in a trap site 
is $\sim50~\mu\rm s$. Both time
constants are much longer than the observed $4~\mu\rm s$ of coherent oscillations. However, the ground-Rydberg phase coherence decays due to Doppler sensitivity of the two-photon excitation according to\cite{note2019} $\langle e^{\imath\phi}\rangle=e^{-t^2/T_{2,\rm D}^2}$ with  $T_{2,\rm D} = \sqrt{2 M_{\rm Cs}/k_B T_a}/k_{2\nu}$ and $k_{2\nu} = 2\pi/\lambda_1-2\pi/\lambda_2.$ At $T_a=15~\mu\rm K$ we find $T_{2,\rm D}=6~\mu\rm s$ which would seem to imply a noticeable decay of the Rabi amplitude.  This is not the case because the effective coherence, or persistence time, of a driven oscillation is longer than that of a static superposition of states \cite{SM2019}.      

The original proposal for a Rydberg $C_Z$ gate\cite{Jaksch2000} involves a sequence of three pulses connecting ground and Rydberg states: a $\pi$ pulse on the control qubit, $2\pi$ on the target qubit, and $\pi$ on the control qubit. It has been shown by detailed analysis of the atomic structure of the heavy alkalis that this pulse sequence is in principle capable of creating entanglement with fidelity $F>0.998$ \cite{XZhang2012}. Many other Rydberg gate protocols have been proposed(see \cite{Saffman2016} for an overview). Using shaped pulses  $F>0.9999$ at gate times 
as short as 50 ns\cite{Theis2016b} appears possible. We report here on improved performance of the original proposal, leaving alternative  protocols for future work. 

\begin{figure}[!t] 
  \includegraphics[width=1.\columnwidth]{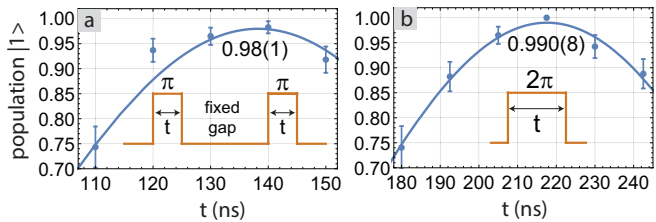}
\vspace{-.7cm}
  \caption{(color online) Population in $\ket{1}$ after   a) control and b) target qubit Rydberg pulses. The Rabi frequencies for the control and target pulses were $\Omega_R/2\pi=3.6$ and $4.6~\rm MHz$ respectively and gap time = 300 ns. }
  \label{fig.pGp}
   \vspace{-.5cm}
\end{figure}

One of the technical errors that has been improved on is the loss of Rydberg atoms  after the $\pi-{\rm gap}-\pi$ pulses on the control qubit or the $2\pi$ pulse on the target 
qubit. These losses dominated the error budget in most earlier experiments\cite{Zhang2010,Wilk2010,Maller2015}. Improvements to  laser noise, optical beam quality, and alignment  have reduced population losses to 2\% as shown in Fig. \ref{fig.pGp}. 
 To minimize excitation of Rydberg hyperfine states with $m_f\ne 0$ we align the $\bf k$ vectors of the 459 and 1038 nm beams to be anti-parallel and set the background magnetic fields and polarization of the 459 nm beam to be accurately $\sigma_+$ relative to a quantization axis along $\bf k$. This is done  by preparing the state $\ket{4,4}$ and minimizing the scattering rate due to the 459 nm light. For the data in Figs. \ref{fig.pGp} - \ref{fig.Bell}  the beam waists were reduced   to $w_1=2.25~\mu\rm m$,  $w_2=2.5~\mu\rm m$  to minimize crosstalk between sites.

The next step in tuning the gate sequence is to verify the qubit phase induced by a $2\pi$ Rydberg pulse. We do this using a Ramsey sequence of $\pi/2$ - gap - $(\pi/2)_\theta$ pulses on the qubit states (the last pulse is about an axis rotated by $\theta$ in the equatorial plane) and insert  a $ 2\pi$ $\ket{1} - \ket{R}$ pulse on the target qubit inside the gap as shown in Fig. \ref{fig.eye}. Performing this sequence with and without first exciting the control qubit to $\ket{R}$ with a $\pi$ pulse gives an ``eye" diagram that ideally consists of blockade and no-blockade curves that are 
$\pi$ out of phase with each other. In the experiment these curves have a relative phase that is not equal to $\pi$ due to Stark shifts of the qubit states induced by the Rydberg excitation beams\cite{Maller2015}. To compensate for this we slightly detune the Rydberg pulse on the target
 to give the opposite phase Ramsey curves seen in the figure.

The observed amplitude of the blockade and no-blockade curves in Fig. \ref{fig.eye} is 0.91(6) and 0.85(3). To prepare a maximally entangled Bell state with the $C_Z$ gate the input state is 
$(\ket{00}+\ket{10}+\ket{01}+\ket{11})/2.$ The $\ket{0}$ state is not Rydberg coupled and is not affected by the gate sequence therefore $\ket{00}$ experiences no error, $\ket{10}$ corresponds to Fig. \ref{fig.pGp}a), $\ket{01}$ corresponds to the no-blockade eye diagram curve, and $\ket{11}$ the blockaded eye diagram curve.  The error channels discussed below in connection with Table \ref{tab.budget} limit the amplitude of these curves.

\begin{figure}[!t] 
  \includegraphics[width=.85\columnwidth]{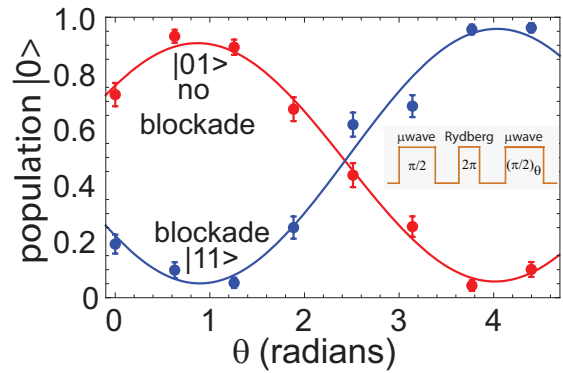}
\vspace{-.4cm}
  \caption{(color online) Eye diagram for target qubit with blockade and no-blockade curves of amplitude 0.91(6) and 0.85(3). The inset shows the target pulse sequence. The $C_Z$ gate was operated at $\theta=0.95$ radians. }
  \label{fig.eye}
   \vspace{-.5cm}
\end{figure}

\begin{table*}[!t]
\caption{Error budget for Bell state preparation using experimental and calculated values.  Individual error sources are combined by propagating the input state through a series of quantum processes using $\chi$-matrices\cite{SM2019}.  
\label{tab.budget}}
\begin{tabular}{l|l|r}
\hline
quantity &  error & fidelity estimate\tabularnewline
\hline\hline
{\bf atomic parameters \& finite temperature effects}  &&\tabularnewline 
 ground-Rydberg Doppler dephasing on control (calculated)  & $0.013$ \tabularnewline
 Rydberg radiative lifetime on control (calculated)  & $0.0075$& \tabularnewline
 scattering $7p_{1/2}$ per atom per $\pi$ pulse per Rydberg beam (calculated)   & $0.0012$& \tabularnewline
 blockade leakage (calculated)   & $0.001$& \tabularnewline
 atom position in Rydberg beams per atom per Rydberg $\pi$ pulse (measured) &  $0.0025$& \tabularnewline
 laser noise per atom per $\pi$ pulse (measured) &  $0.0025$& \tabularnewline
 Rydberg crosstalk for $\ket{01}$ (measured) &  $0.005$& \tabularnewline
Rydberg laser dephasing per atom per Rydberg $\pi$ pulse (measured) &   $0.018$ (nonblockaded)& \tabularnewline
    &  $0.006$ (blockaded) & $F_{\rm Bell}^{C_Z}=0.887$\tabularnewline
\hline 
{\bf single qubit errors}  & & \tabularnewline
 global per atom per $\mu$wave $\pi/2$ pulses (4 total) (measured)  &  $0.0028$&   \tabularnewline
 Stark-shift pulse (measured)  &  $0.006$& $F_{\rm Bell}^{\rm -SPAM}=0.877$\tabularnewline
\hline 
{\bf SPAM errors}  & &\tabularnewline
 readout loss per atom per readout (initial and final) (measured)  & $0.0025$ &  \tabularnewline
 optical pumping per atom (estimated) & $0.005$ &  \tabularnewline
 state measurement error per atom (measured)  & $0.00015$ &  $F_{\rm Bell}=0.853$\tabularnewline
\hline
\end{tabular}
\vspace{-.3cm}
\end{table*}

To prepare a Bell state we initialize the entire array, including control and target qubits, in state $\ket{1}$ and  post-select on cases when both target and control sites are filled.  We do not post-select on occupation of any of the neighboring sites so the results are an average over a mean of 55\% occupation. We then perform the pulse sequence shown in Fig.\ref{fig.Bell} which puts the control qubit in a superposition of $\ket{0}$ and $\ket{1}$ and implements a CNOT gate using a combination of the Rydberg $C_Z$ and $\pi/2$ rotations on the target qubit.  To simplify the pulse sequence, the initial $\pi/2$ rotations were performed with a global microwave pulse.  After the $C_Z$, we perform  a  $\pi/2$ rotation of the target qubit alone  about an axis $\theta$ to complete the CNOT gate\cite{SM2019}.

The populations of the two-qubit output state are measured, and the coherence is determined from the amplitude of parity oscillations due to a global microwave $\pi/2$ rotation at variable angle $\phi$\cite{Maller2015}. The resulting data shown in Fig. \ref{fig.Bell} gives $(P_{00}+P_{11})/2=0.47(2)$, parity amplitude $C=0.391(6)$, and $F_{\rm Bell}= 0.47 + 0.39 = 0.86(2)$. Note that since a global $\pi/2$ microwave pulse is used to prepare the superposition $(\ket{00}+\ket{10}+\ket{01}+\ket{11})/2$, all atoms in the array occupy an equal superposition of $\ket{0}$ and $\ket{1}$ during the $C_Z$ gate sequence.  This superposition is representative of the average situation expected when performing a larger  quantum algorithm. 

\begin{figure}[!t] 
  \includegraphics[width=.95\columnwidth]{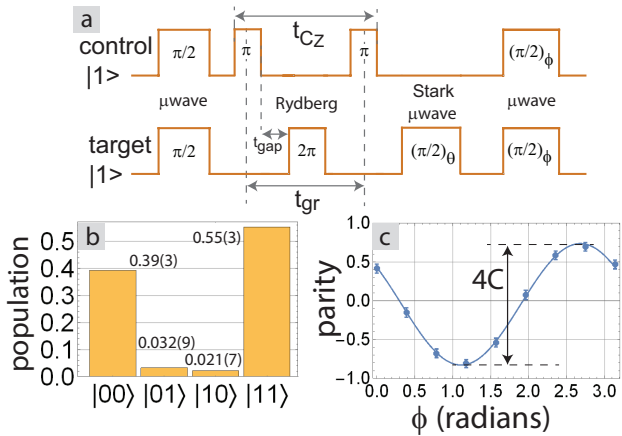}
\vspace{-.4cm}
  \caption{(color online) Bell state preparation: a) pulse sequence, b) populations, c) parity oscillation. The Rydberg pulses had lengths of $t_\pi=150~\rm ns$, $t_{2\pi}=220~\rm ns$, $t_{\rm gap}=300~\rm ns$. The effective ground-Rydberg superposition time is $t_{\rm gR}=t_\pi/2+t_{\rm gap}+t_{2\pi}+t_{\rm gap}+t_{\pi}/2=0.98~\mu\rm s.$ The microwave pulses are global (duration $35~\mu\rm s$) for the initial and parity pulses and global pulses combined with a 459 nm Stark pulse (duration $70~\mu\rm s$) for the site selected rotation.}
  \label{fig.Bell} \vspace{-.9cm}
\end{figure}

The observed Bell fidelity can be understood from the error sources 
enumerated in Table \ref{tab.budget}. The errors are divided into three categories: a) errors due to atomic parameters, finite temperature, laser noise, and crosstalk between sites, b) errors in the single qubit operations used for the CNOT gate and parity measurement, and 
c) SPAM errors. Calculations and measurements supporting the error model are provided in the supplementary material\cite{SM2019}.   The effect of each of these errors on the output Bell state is  calculated by modeling each error source as a quantum process.  By propagating the input state through each of these quantum processes, we are able to accurately account for correlations between errors and the total resulting infidelity of the output  Bell state \cite{SM2019}.   
 Including all errors we arrive at the value in the last line of Table \ref{tab.budget} which is consistent with the observed
$F_{\rm Bell}= 0.86(2).$ Accounting for SPAM errors gives a corrected Bell fidelity $F_{\rm Bell}^{-\rm SPAM}= 0.88.$ Accounting for single qubit errors that go into the CNOT gate and parity measurements we arrive at a SPAM and single qubit gate corrected fidelity of $F_{\rm Bell}^{\rm C_Z}= 0.89.$

Since the array filling was not deterministic, the crosstalk from neighboring sites is approximately half of what it would be in a fully occupied array.  Using our error model\cite{SM2019}, we estimate that in a deterministically loaded array we would see a reduction in Bell state fidelity of about $0.4\%$.  Figure \ref{fig.array}d shows that readout retention varies over the array with a median  error that is $0.8\%$ per atom per readout higher than the sites selected in this experiment.  Using the error model\cite{SM2019} we estimate that the median observed Bell state fidelity for neighboring sites would be about $2.1\%$ lower than for the sites that were used here.

The Table shows that the dominant $C_Z$ gate errors are due to finite temperature which leads to atomic motion and dephasing, and atomic position variations in the optical traps. In addition laser intensity and phase noise contribute to the gate error at the percent level. Errors due to the finite radiative lifetime of excited states and the finite blockade strength contribute less than 1\%. These observations support the potential for  Rydberg  gates that have fidelity compatible with fault tolerant error correction after further technical improvements for reduced atom temperature and laser noise reduction.

In summary we have used a two-qubit Rydberg $C_Z$ gate to create a Bell state with intrinsic fidelity after correcting for SPAM and single qubit errors of $F_{\rm Bell}^{C_Z}=0.89.$
This improved  fidelity was obtained in a 2D qubit array  using  tightly focused control beams scanned with acousto-optic deflectors making site specific gate operations across the array possible.  Therefore, the existing experimental setup allows gates on a number of sites throughout the array and provides an architecture for scaling to large quantum algorithms. 

During completion of our manuscript we became aware of related work demonstrating parallel operation of Rydberg gates\cite{Levine2019}.  

We acknowledge support from NSF PHY-1720220, the ARL-CDQI, DOE award DE-SC0019465, and ColdQuanta, Inc. . 


%

\pagebreak

{. }

\newpage

\newpage

\begin{widetext}

\setcounter{figure}{0}
\renewcommand\thefigure{SM-\arabic{figure}}
\setcounter{table}{0}
\renewcommand\thetable{SM-\Roman{table}}

\setcounter{section}{0}
\renewcommand\thesection{SM-\Roman{section}}

\section*{Supplementary Material for\\ Rydberg mediated  entanglement in a two-dimensional neutral atom qubit array}

\title{Rydberg mediated entanglement in a two-dimensional neutral atom qubit array} 

\author{T. M. Graham, M. Kwon, B. Grinkemeyer, Z. Marra, X. Jiang}
\author{M. T. Lichtman}
\altaffiliation{Present address: 
Joint Quantum Institute and Department of Physics,
University of Maryland, College Park, MD 20742}
\author{Y. Sun}
\altaffiliation{Present address: Interdisciplinary Center for Quantum Information,
National University of Defense Technology, Changsha 410073, China}
\author{M. Ebert}
\altaffiliation{ColdQuanta, Inc., 811 E. Washington Ave, Suite 408, Madison, WI 53703}
\author{M. Saffman}
\altaffiliation{ColdQuanta, Inc., 811 E. Washington Ave, Suite 408, Madison, WI 53703}
\affiliation{
Department of Physics, University of Wisconsin-Madison, 1150 University Avenue, Madison, Wisconsin 53706
}

\date{\today}
\maketitle

\setcounter{figure}{0}
\renewcommand\thefigure{SM-\arabic{figure}}
\setcounter{table}{0}
\renewcommand\thetable{SM-\Roman{table}}
\setcounter{section}{0}
\renewcommand\thesection{SM-\Roman{section}}

\tableofcontents
\vspace{1.cm}

In this supplementary material we reproduce Table I from the main text and provide explanations, additional measurements,  and underlying calculations that support the error model. The values of the assigned errors are explained in Secs. \ref{sec.ErrorCalc}, \ref{sec.singleQubitErrors}, \ref{sec.SPAMErrors}. The error modeling builds on earlier analyses\cite{Saffman2005a,Saffman2010,Saffman2011,Zhang2010,XZhang2012,Gillen-Christandl2016} supplemented by new material concerning atom position variations with respect to the Rydberg beam envelopes, and the influence of laser phase noise on gate fidelity. The combined effect of the error channels listed in Table \ref{tab.budget2} is calculated using process matrices as described in Secs. \ref{sec.ErrorProp}, \ref{sec.processmatrix}. Section \ref{sec.rabiramsey} explains the statement in the main text that  the observed coherence time of a driven Rabi oscillation is longer than the measured Ramsey coherence time. 
 
\newpage

\begin{table*}[!t]
\caption{Error budget for Bell state preparation using experimental and calculated values. The listed error values are calculated in Sections \ref{sec.ErrorCalc} -- \ref{sec.SPAMErrors}.  Individual error sources are combined by propagating the input state through a series of quantum processes using $\chi$-matrices as described in Sec. \ref{sec.ErrorProp}.  This is the same as Table I in the main text.
\label{tab.budget2}}
\begin{tabular}{l|l|r}
\hline
quantity &  error & fidelity estimate\tabularnewline
\hline\hline
{\bf a) atomic parameters \& finite temperature effects}  &&\tabularnewline 
a.1) ground-Rydberg Doppler dephasing on control (calculated)  & $0.013$ \tabularnewline
a.2) Rydberg radiative lifetime on control (calculated)  & $0.0075$& \tabularnewline
a.3) scattering $7p_{1/2}$ per atom per $\pi$ pulse per Rydberg beam (calculated)   & $0.0012$& \tabularnewline
a.4) blockade leakage (calculated)   & $0.001$& \tabularnewline
a.5) atom position in Rydberg beams per atom per Rydberg $\pi$ pulse (measured) &  $0.0025$& \tabularnewline
a.6) laser noise per atom per $\pi$ pulse (measured) &  $0.0025$& \tabularnewline
a.7) Rydberg crosstalk for $\ket{01}$ (measured) &  $0.005$& \tabularnewline
a.8) Rydberg laser dephasing per atom per Rydberg $\pi$ pulse (measured) &   $0.018$ (nonblockaded)& \tabularnewline
    &  $0.006$ (blockaded) & \tabularnewline
& { } &$F_{\rm Bell}^{C_Z}=0.887$\tabularnewline
\hline 
{\bf b) single qubit errors}  & & \tabularnewline
b.1) global per atom per $\mu$wave $\pi/2$ pulses (4 total) (measured)  &  $0.0028$&   \tabularnewline
b.2) Stark-shift pulse (measured)  &  $0.006$&   \tabularnewline
&{ }&$F_{\rm Bell}^{\rm -SPAM}=0.877$\tabularnewline
\hline 
{\bf c) SPAM errors}  & &\tabularnewline
c.1) readout loss per atom per readout (initial and final) (measured)  & $0.0025$ &  \tabularnewline
c.2) optical pumping per atom (estimated) & $0.005$ &  \tabularnewline
c.3) state measurement error per atom (measured)  & $0.00015$ &  \tabularnewline
&{ }&$F_{\rm Bell}=0.853$\tabularnewline
\hline
\end{tabular}
\end{table*}

\section{Experimental setup, atomic parameters, and finite temperature effects}
\label{sec.ErrorCalc}

In this and the following two sections, we identify  error contributions which influence the Bell state preparation and measurement described in the main manuscript.  We then calculate, or estimate based on measurements, the magnitude of each of these effects.  The calculation of how these errors collectively influence the measured Bell state is performed in Sec. Secs. \ref{sec.ErrorProp} using a $\chi$-matrix analysis.

The experimental setup is an upgraded version of that used in \cite{Maller2015}.  The central part of the experimental arrangement is shown in Fig. \ref{fig.SM setup}. The primary differences relative to \cite{Maller2015} are a new method of creating the trap array\cite{Lichtman2015,Graham2019b}, an upgraded all-glass UHV cell that includes internal electrodes for compensating background electric fields, higher numerical aperture objective lenses, and improvements to  locking electronics for reduced noise on the Rydberg excitation lasers. We proceed to discuss various sources of experimental imperfection. 

\subsection{a.1 Ground-Rydberg Doppler dephasing }
\label{subsec.Doppler}
Rydberg excitation is performed with counterpropagating beams  of different wavelengths so there is a  wavenumber mismatch $k_{2\nu}=2\pi/\lambda_1-2\pi/\lambda_2$. This leads to a stochastic phase for atoms that are Rydberg excited, spend a time $t_{\rm gR}$ in the Rydberg state, and then de-excited. This was originally pointed out in \cite{Wilk2010}, although our analysis\cite{note2019,Saffman2011} gives a  smaller prediction for the magnitude of the phase error.

The average stochastic phase term can be expressed as $\langle e^{\imath\phi} \rangle=e^{-t_{\rm gR}^2/T_{2,\rm D}^2}$ with $T_{2,\rm D} = \sqrt{2 M_{\rm Cs}/k_B T_a}/k_{2\nu}$. 
At the measured atomic temperature of $T_a=15~\mu\rm K$ we find $T_{2,\rm D}=6~\mu\rm s$. This effect limits the Bell 
4
 fidelity to $F_{\rm Bell}^{\rm max}=(1+\langle e^{\imath\phi} \rangle)/2=0.987$ using $t_{\rm gR}=0.98~\mu\rm s$. Our measurement of the ground-Rydberg coherence decay using a Ramsey experiment reveals a shorter coherence time of $T_{2,\rm gR}=4~\mu\rm s$ which would imply  
$F_{\rm Bell}^{\rm max}=0.971$. We do not use this value in the error table because the difference between $T_{2,\rm D}$ and $T_{2,\rm gR}$ is assumed due to other error mechanisms, including Rydberg radiative lifetime, atom position variations,  and laser noise, that are separately included in the table.

\subsection{a.2 Rydberg radiative lifetime }
\label{subsec.Lifetime}

The lifetime of the Cs $66s$ state in a room temperature bath is calculated to be \cite{Beterov2009,*Beterov2009b}
$\tau = 130~\mu\rm s$. Input state $\ket{ij}$ will experience a  spontaneous emission error of $\epsilon_{ij}=1- e^{-t_{ij}/\tau}$ with $t_{ij}$ the Rydberg population integrated over the duration of the $C_Z$ gate sequence for the corresponding input state.   Thus $t_{00}=0$ since $\ket{00}$ is not Rydberg coupled, $t_{01}=t_{2\pi}/2$ for state $\ket{01}$,  $t_{10}=t_{11}=t_{\rm gR}$ for states $\ket{10},\ket{11}$. Using $t_{2\pi}=0.22~\mu\rm s$ and $t_{\rm gR}=0.98 ~\mu\rm s$ we find 
$\epsilon_{00}=0, \epsilon_{01}=8.5\times 10^{-4}, \epsilon_{10}=\epsilon_{11}=7.5\times 10^{-3}$. This error is less than 1\% even for the relatively low lying $n=66$ state used here.

\begin{figure}[!t]
    \centering
    \includegraphics[width=.9\columnwidth]{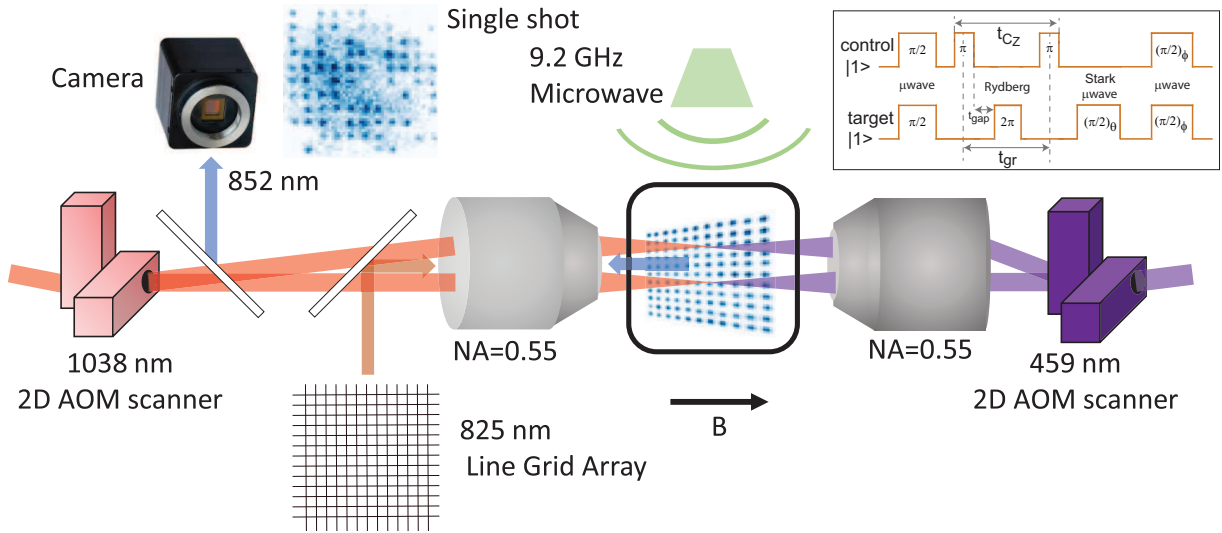}
    \caption{Schematic of the experimental geometry. The atomic array is located in an all glass UHV cell. Trap light at 825 nm and qubit control beams at 459 and 1038 nm are projected through facing objective lenses. Qubit measurements use scattered light at 852 nm that is imaged onto a EMCCD camera. A microwave horn is used for single qubit gate operations. The inset shows the experimental pulse sequence. }
    \label{fig.SM setup}
\end{figure}

\subsection{a.3 Scattering from $7p_{1/2}$  }

Two-photon excitation is via the $7p_{1/2}$ state which has a radiative lifetime of $\tau_{7p}=155 ~\rm ns$. The one-photon detuning is 
$\Delta/2\pi = 680~\rm MHz$. The probability of spontaneous emission in a $\pi$ pulse when the one-photon Rabi frequencies are equal is $P_{\rm se}=\frac{\pi}{2}\frac{1}{\tau_{7p}\Delta}=0.0023.$ When the Rydberg excitation is blockaded, only light from the 459 nm Rydberg laser contributes to scattering from  $7p_{1/2}$, so the scattering probability in a $\pi$ pulse is halved, yielding $P_{\rm se}=\frac{\pi}{4}\frac{1}{\tau_{7p}\Delta}=0.0012.$   Here we are neglecting corrections due to the hyperfine structure of the $7p_{1/2}$ state. The full expressions, including those corrections, can be found in \cite{Maller2015}.

\subsection{a.4 Blockade leakage  }

Due to less than infinite blockade strength there is finite rotation of the blockaded state. The average error is \cite{Saffman2010}
$
\epsilon= \Omega_R^2/8 {\sf B}^2.$
The calculated blockade strength for Cs $66s$ at a distance of 
$R=\sqrt2 d$ is ${\sf B}/2\pi= 45~\rm MHz$ which gives $\epsilon = 0.001$.  Measurements of the blockade showed less than $0.02$ leakage which is consistent with SPAM estimates given below. 

\subsection{a.5 Atom position in Rydberg beams }
\label{subsec.Spatial}
All the reported measurement results are averaged over multiple realizations and for each realization the atomic position is slightly different. Averaging over atomic positions leads to decay of coherent oscillations, an effect that was studied numerically in \cite{Gillen-Christandl2016}.
Here we calculate the expected errors and their uncertainties for $\pi$ and $2\pi$ pulses. This contribution to the error budget, as well as the contribution from laser noise in the following section, has larger uncertainty then the other errors due to lack of precise knowledge of the optical
parameters of the array. In particular the width and shape of the lines defining the trapping sites affect the atomic localization but are only known approximately.   

 The position distribution of a trapped atom at finite temperature can be modeled as a Gaussian 
with radial and axial localization parameters $\sigma,\sigma_z$
and normalized distribution
$$
\rho({\bf r})=\frac{1}{(2\pi)^{3/2}\sigma^2\sigma_z}e^{-(x^2+y^2)/(2\sigma^2)}e^{-z^2/(2\sigma_z^2)}.
$$
The Rydberg beam field amplitudes relative to their values at the center of the beam are 
$$
f_j({\bf r})=\frac{e^{-(x^2+y^2)/w_j^2(z)}}{\sqrt{1+z^2/L_{R_j}^2}}.
$$
Here $w_j^2(z)=w_j^2(1+z^2/L_{R_j}^2)$ are the $z$ dependent beam sizes, $L_{R_j}=\pi w_j^2/\lambda_j$ are the Rayleigh lengths, and $j=1,2$ refer to two Rydberg excitation beams. We assume that the Rydberg beams are aligned with the origin of the atom distribution. 
If $\Omega_R$ is the two-photon Rabi frequency at the center of the beams then the position dependent Rabi frequency is 
$\Omega_R({\bf r})=\Omega_R f_1({\bf r})f_2({\bf r}).$

Consider a $2\pi$ pulse of length $t$ with the initial state $\ket{i}$. The average observed probability for the atom to be in $\ket{1}$ after the pulse  is 

\begin{eqnarray}
\langle P_{\ket{i}}(t)\rangle
=\int d{\bf r}\, \rho({\bf r})\left\{\cos^2\left[ \frac{\Omega_R t}{2}f({\bf r})\right]+\frac{\Delta^2({\bf r})}{\Omega_R^2({\bf r})+ \Delta^2({\bf r})} \sin^2\left[ \frac{\Omega_R t}{2}f({\bf r})\right] \right\}.
\label{eq.pop}
\end{eqnarray}

This expression includes the position dependent two-photon detuning 
$\Delta({\bf r})=\Delta_0 + \Delta_1f_1^2({\bf r})
+ \Delta_2f_2^2({\bf r}).$ Here $\Delta_0$ is a constant detuning and $\Delta_1, \Delta_2$ are Stark shift coefficients that depend on the beam intensities and detunings relative to the intermediate $7p_{1/2}$ state. 
Choosing $\Delta_0=-(\Delta_1+\Delta_2)$ ensures that an atom at trap center is resonantly excited.

The variance of the  population error is
\begin{eqnarray}
\left(\delta P\right)^2 &=& \left\langle\left[ P_{\ket{i}}(t)-P_{\rm tar}\right]^2 \right\rangle\nonumber\\
&=& P_{\rm tar}^2 + \left\langle \left[ P_{\ket{i}}(t)\right]^2\right\rangle -2 P_{\rm tar}\langle P_{\ket{i}}(t)\rangle .
\label{eq.var}
\end{eqnarray}

The pulse time $t$ can be adjusted to minimize the difference between the observed average $\langle P_{\ket{i}}(t)\rangle$ and 
the target population $P_{\rm tar}$. For a $2\pi$ pulse 
$P_{\rm tar}=1$  but, at finite temperature,  there is no pulse time for which $\langle P_{\ket{i}}(t)\rangle=P_{\rm tar}.$ Nevertheless the time can be adjusted to minimize the difference between the target population and the observed average. For a $2\pi$ pulse 
\begin{eqnarray}
\left(\delta P\right)^2 
&=& 1 + \left\langle \left[ P_{\ket{i}}(t)\right]^2\right\rangle -2 \langle P_{\ket{i}}(t)\rangle .
\end{eqnarray}

Similar expressions govern the phase response. 
The amplitude of the initial state after a pulse of length $t$ is
\begin{eqnarray}
c({\bf r},t)&=& \cos\left[\sqrt{\Omega_R^2({\bf r})+\Delta^2({\bf r})}t/2\right]\nonumber\\
 &-& i \frac{\Delta({\bf r})}{\sqrt{\Omega_R^2({\bf r})+\Delta^2({\bf r})}}\sin\left[\sqrt{\Omega_R^2({\bf r})+\Delta^2({\bf r})}t/2\right]\nonumber\\
&&
\label{eq.phase}
\end{eqnarray}

The phase of the wavefunction is 
\begin{equation}
\phi=-\tan^{-1}\left[\frac{\Delta({\bf r})}{\sqrt{\Omega_R^2({\bf r})+\Delta_R^2({\bf r})}}\tan\left[\frac{\sqrt{\Omega_R^2({\bf r})+\Delta^2({\bf r})}}{2}t\right] \right].
\label{eq.phase2}\end{equation}
Expressing $\Omega_R({\bf r})$ and $\Delta({\bf r})$ in terms of $f_1, f_2$ and integrating over $\rho({\bf r})$ as for the  population distribution we calculate the phase error and uncertainty.

Perturbative analysis of the population variance in 2D for $\sigma\ll w$
shows that $(\delta P)^2\sim (\sigma/w)^8$ so the standard deviation or uncertainty in the population scales as 
$(\sigma/w)^4$. This scaling highlights the sensitivity to finite beam size and the importance of low temperatures and 
tight localization of the atom. Sensitivity to variations 
in the local Rabi frequency and detunings can be minimized using adiabatic gate protocols (see, for example, \cite{Beterov2013a}), but we have not done so in the experiments reported here.

\end{widetext}

To determine the errors from atom position variations in 3D we have relied on numerical solutions of Eqs. ({\ref{eq.pop}-\ref{eq.phase2}). The atom localization parameters $\sigma, \sigma_z$ were estimated\cite{Graham2019b} using $T_a=15~\mu\rm K$, trap depth 
$285~\mu\rm K$, $d=3.1~\mu\rm m$, and line width $w_{\rm line}=d/3.08=1.0~\mu\rm m$ to be 
$\sigma=0.27~\mu\rm m$, 
$\sigma_z=1.47~\mu\rm m$. 
The resulting pulse errors deduced from  numerical calculations are shown in Fig. \ref{fig.popvar}. The population error corresponding to the values of $\sigma, \sigma_z$ deduced from optical parameters  is too large to be consistent with the observed population after a ground-Rydberg Rabi pulse shown in Fig. 3 in the main text, while the phase 
error as quantified by $\langle 1-\cos(\phi)\rangle$ is negligible. We attribute the discrepancy to uncertainties in the optical parameters in part due to aperturing of the array light pattern on the objective lens aperture which caused broadening of the lines and reduced positional spread. The  observed trapped atom vibrational frequencies also imply tighter confinement.  Accounting for the other  error sources that contribute to the Rabi pulse population error our best estimate for the transverse localization is $\sigma=0.16~\mu\rm m$ which is indicated by a vertical line in the figure and  implies a $2\pi$ pulse population error of 0.006.

\begin{figure*}[!t] 
  \includegraphics[width=1.9\columnwidth]{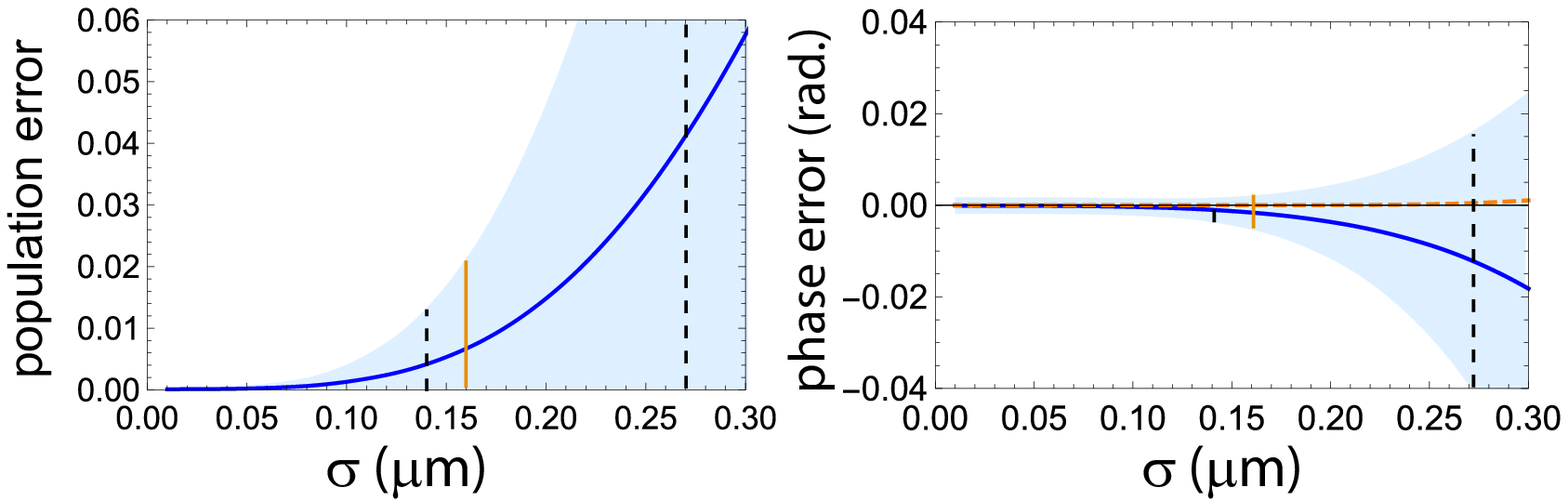}
  \caption{Numerical solutions for the population and phase errors due to atom position variations after a $2\pi$ ground-Rydberg pulse. a) Population error with $\pm$ one standard deviation shown by the light blue shading. b) Phase error with $\langle\phi\rangle$ shown by the solid blue line, $\langle 1-\cos(\phi)\rangle$ shown by the dashed orange line, and the standard deviation shown by the light blue shading. The vertical dashed lines indicate the range within which $\sigma$ is bounded based on optical trapping parameters (maximum value of $\sigma$) and consistency with observed errors in ground-Rydberg Rabi rotations (minimum value of $\sigma$). The vertical orange line indicates the value of $\sigma$ used for the gate error analysis.  
Parameters:  $w_1=2.25~\mu\rm m$,
$w_2=2.5~\mu\rm m$, $\Omega_R/2\pi=4.5~\rm MHz$, $\Delta_1/2\pi=-2.7~\rm MHz$, $\Delta_2/2\pi=6.4~\rm MHz$, $\Delta_0=-(\Delta_1+\Delta_2).$
}
\label{fig.popvar}
\end{figure*}

The $\pi$ Rydberg pulse on the control atom also contributes an additional error since less than 100\% population transfer from ground to Rydberg state implies a corresponding failure probability for blockade of the target qubit in the $\ket{11}$ state. The probability of populating the Rydberg state after a $\pi$ pulse is 
\begin{eqnarray}
\langle P_{\ket{R}}(t)\rangle
=\int d{\bf r}\, \rho({\bf r}) \frac{\Omega_R^2({\bf r})}{\Omega_R^2({\bf r})+ \Delta^2({\bf r})} \sin^2\left[ \frac{\Omega_R t}{2}f({\bf r})  \right].
\label{eq.popvar2}
\end{eqnarray}

The calculated population error for a $\pi$ pulse using Eq. (\ref{eq.popvar2}) is shown in Fig. \ref{fig.popvar2}.  The error contribution at $\sigma=0.16~\mu\rm m$ is $\epsilon=0.0025$.

\begin{figure}[!t] 
  \includegraphics[width=.9\columnwidth]{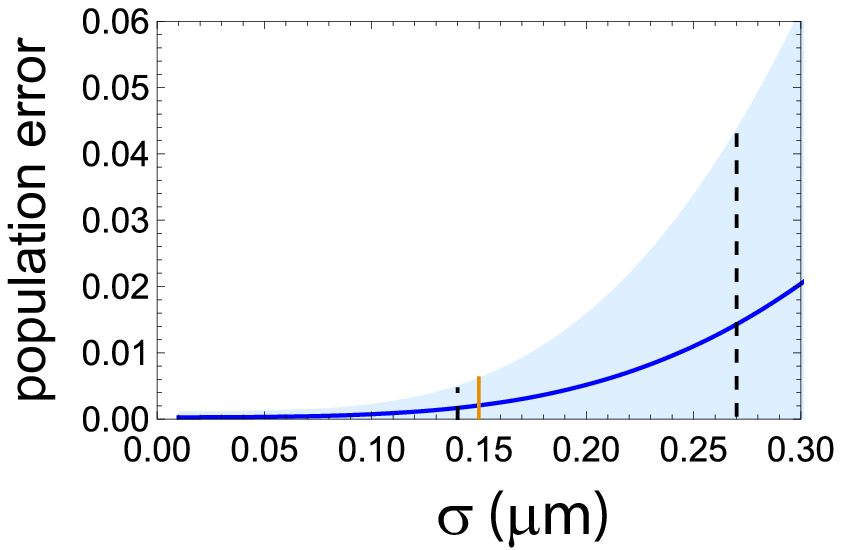}
  \caption{Numerical solutions for the population   error  due to atom position variations after a $\pi$ ground-Rydberg pulse with $\pm$ one standard deviation shown by the light blue shading.    All other 
parameters  the same as in Fig.  \ref{fig.popvar}. 
}
\label{fig.popvar2}
\end{figure}

We emphasize that our confidence in this error estimate is lower than for other entries in the error model, due to lack of precise knowledge of trap parameters and atom localization. Furthermore these errors will be larger if the Rydberg beams are misaligned relative to the trap centers. There is also a correlation with the Rydberg atom motion (see Sec. \ref{sec.othererrors}) whereby a Rydberg excited atom will move and see a different beam amplitude under deexcitation. We have not included such effects in the analysis.

\subsection{a.6 Laser noise}
\label{subsec.LaserNoise}
Laser noise due to intensity and phase fluctuations degrades the fidelity of coherent pulses. Intensity noise causes undesired Stark shifts, analogous to the effects of atom motion. The linewidths of the Rydberg lasers determined by beating together two lasers built using the same components were determined to be under 300 Hz. However, locking to 5 kHz linewidth stabilized reference cavities leads to ``servo bumps" and phase noise which can degrade the fidelity of Rabi pulse operations\cite{deLeseleuc2018}.

The pulse errors can be significantly reduced by resonator filtering\cite{Levine2018}. While we have demonstrated persistent coherence of ground-Rydberg oscillations in Fig. 2 of the main text without resonator filtering, it is also the case that significant day to day variations in Rabi oscillation amplitude were observed that could not always be directly correlated with the observable amplitude of servo bumps on the lasers. 

Simulations we have performed by modeling servo bump noise as a superposition of frequency components with random phases confirm that the pulse errors are largest when the servo bump offset frequency is comparable to $\Omega_R$. For integrated servo bump power of $-20~\rm dBr$ relative to the carrier we find population errors up to 0.04, and for $-30~\rm dBr$
errors up to 0.002. Temperature and current drifts in laser diodes result in changes in the laser mode  structure and noise levels that also affect the experimental results, even without changes to lock parameters. This error is therefore difficult to quantify, but is certainly present since we have observed large changes in pulse errors with minor changes to the locking electronics parameters. 
A reasonable estimate for this error is in between the 
$-20~\rm dBr$ and $-30~\rm dBr$ values given above. We have placed this error at 0.0025 per atom per Rydberg $\pi$ pulse in the error model, but there are relatively large uncertainties. 

\subsection{a.7 Rydberg crosstalk}
Another source of error arises from the presence of other atoms in the optical lattice.  Specifically, Rydberg excitation of non-addressed sites by Rydberg beam intensity spilling over to neighboring sites can influence the state of control and target atoms.  It should be noted that this error primarily arises during the first Rydberg $\pi$ pulse on the control atom.  During the first pulse, if a control atom is in the $\ket{0}$ state, there is a possibility for one of the control atom's neighbors to be excited to the Rydberg state (if the control atom is in the $\ket{1}$ state, Rydberg blockade greatly suppresses the excitation of neighboring sites).  When a neighboring atom is excited, the target atom is erroneously blockaded.  We determined from Rydberg $\pi$ pulse experiments that there is $\leq0.0025$ chance that a neighboring atom in $\ket{1}$ will be excited to the Rydberg state.  There are eight neighboring sites to the control site, including the target site. These sites are randomly filled and in an equal superposition of $\ket{0}$ and $\ket{1}$.  This implies that there is a cross-talk error probability $\leq0.005$.  Note, for a fully occupied array, this error would be almost a factor of two larger. 

We can estimate expected crosstalk levels from beam parameters and the array period.
The nominal Rabi frequency at  neighboring sites due to Rydberg beam crosstalk is  $\Omega_R'=e^{- d^2/w_1^2}e^{- d^2/w_2^2}\Omega_R=0.12~ \Omega_R$, while the data in the main text Fig. 2 shows no evidence of Rabi oscillations at the other sites.  We achieve reduced crosstalk by using beam powers that give a nonzero differential Stark shift of $\Delta'/2\pi= 2~\rm MHz$ between ground and Rydberg states which suppresses the oscillation amplitude to $(\Omega_R'/\Delta')^2=0.023$. These types of errors can be further reduced by using super-Gaussian Rydberg beams \cite{Gillen-Christandl2016}. 

\subsection{a.8 Rydberg laser dephasing}
In addition to the errors described above, we measure an additional dephasing error between ground and Rydberg states arising when an atom is illuminated by Rydberg lasers.  We have measured this error directly through ground-Rydberg Ramsey measurements where one of the two Rydberg lasers was pulsed on the atom during the gap between Rydberg $\pi/2$ pulses.  In addition, we can directly measure the contribution of this error in eye diagram measurements as described in the main text.  Details of how we characterized Rydberg laser dephasing are listed below in subsection \ref{subsec.CZErrorProp}.  The error probability of this effect is $0.018$ per atom per Rydberg $\pi$ pulse and $0.006$ per atom per blockaded Rydberg $\pi$ pulse.

\subsection{a.9 Additional errors}
\label{sec.othererrors}

There are several additional sources of error that are not included in the Table but will be relevant for future, higher fidelity gate demonstrations. 

The first is the magnetic sensitivity of Rydberg states. Rydberg excitation is performed at a bias field of 0.6 mT which is deep in the hyperfine  Paschen-Back limit for the Cs $66s_{1/2}$ state. Therefore the $m_j=-1/2$ state has a magnetic sensitivity of approximately 14. GHz/T. The experiment is operated with synchronization to the 60 Hz AC power line to minimize magnetic noise. Although we have not carefully characterized magnetic noise at the location of the atoms, on the basis of measured atomic temperature and coherence times we estimate that the characteristic magnetic noise amplitude is  $\sigma_B<10^{-6}~\rm T$. With the assumption of Gaussian statistics for the noise the effective 
coherence time is \cite{Saffman2011}
$$
T_{2B}=\frac{2^{3/2}\pi\hbar}{|g_{\rm R}m_{j\rm R}-g_gm_{f\rm g}|\mu_B\sigma_B}.
$$
With the small magnetic bias field used here the ground state magnetic sensitivity is negligible, and using 
$g_{\rm R}m_{j\rm R}=1$ for the Rydberg Zeeman state we find $T_{2B}=0.1~\rm ms$. This implies a coherence loss of $1-e^{-t_{\rm gR}^2/T_{2B}^2}=0.0001$ which is negligible compared to other errors.

The second additional error is motional loss of atoms since we turn the optical traps off during the $C_Z$ gate. 
 If the atom moves outside of the trapping region before the trap light is turned on again it will be lost, leading to an error.  Even if it does not move completely out of the trapping region, it will sit at a higher potential energy when the trap is turned back on.  If an atom gains enough energy, it can escape from the trap, even if it did not move completely outside of the trapping region.  To calculate the magnitude of this effect, we first determine the escape velocity of our trap given its size, depth, and the ratio of axial to transverse trap depth (because the transverse trap depth is about 1.4 times deeper than the axial trap depth, the trap depth is limited by the axial depth).  The atom motion is three dimensional, but the distance for trap loss is much longer in the axial direction, so we approximate loss as being dominated by transverse atom motion.  The escape velocity, $v_e$, is then the transverse velocity, $v_\perp$, that it takes to travel $1.2~ \mu \rm m$ (the point at which the atom would gain enough potential energy to escape once the trap is turned back on) during the trap drop time.  Once the $v_e$ is determined, we find the probability that the atom escapes, $P_{\rm escape}$, by integrating the Maxwell-Boltzmann distribution for transverse velocity, $f({v_\perp})$, from $v_e$ to infinity

\begin{eqnarray}
P_{\rm escape} =\int_{v_{e}}^{\infty} dv_\perp \, v_\perp  f({v_\perp})\, =\frac{m}{k_B T} \int_{v_{e}}^{\infty} dv_\perp\, v_\perp e^{-\frac{m v_{\perp}^2}{2 k_B T}}  \, ,
\end{eqnarray}
where $m$ is the mass of a Cs atom, $T$ is the measured atom temperature, and $k_B$ is the Boltzmann constant.  After performing this integration, we find that the escape probability during the $1.7~ \mu \rm s$ trap drop period of the $C_Z$ gate is orders of magnitude smaller than other errors included in our model.  This result was confirmed using a Monte-Carlo simulation of trap-drop and recapture for our trapping potential using finite atom localization in the trap.

An additional potential source of error is the high sensitivity of Rydberg atoms to dc and ac electric fields. We have previously observed strong sensitivity of specific Rydberg states to background microwave fields when there are resonant transitions driven by cell phones or wi-fi networks. For the $66s_{1/2}$ state the lowest frequency transition to a $n'p$ state is at 12.5 GHz and the lowest frequency two-photon transition is 27. GHz for coupling to a $n's$ state, and 
11.7 GHz for coupling to a $n'd$ state. These frequencies are far from any wireless network bands and not expected to cause significant perturbations. 

There is also sensitivity to time varying dc fields that are of concern up to a few MHz in frequency, corresponding to the bandwidth of the Rydberg pulses.  The experiments are performed in glass cells with eight internal electrodes for field cancellation. The cells are manufactured by ColdQuanta, Inc. . The atomic array is 1 cm from the inside of the cell windows. Voltages generated by a low noise dc supply are used to cancel the background dc field as measured by Rydberg spectroscopy. When the field is minimized there is a quadratic maximum of the Rydberg energy and a zero derivative with respect to field strength which minimizes the Rydberg state sensitivity to fluctuations. The fields are checked every few months and significant changes relevant to the zero field condition have not been observed. We cannot presently exclude fluctuating fields that may cause Rydberg dephasing, but since we are unable to make a reliable estimate of the magnitude of this effect we have left  it out of the error table.

\section{Single qubit errors }
\label{sec.singleQubitErrors}

Single qubit rotations are performed with microwaves and focused Stark shifting lasers. These operations give additional errors that contribute to the infidelity of the Bell state and a CNOT gate but are separate from the fidelity of the Rydberg $C_Z$ gate. 

\subsection{b.1 Global microwave pulses}
\label{subsec.uwave}

The observed amplitude  of microwave $\pi$ pulses connecting the clock states is $0.988(4)$ for the control site and $0.985(4)$ for the target site 
is dependent on microwave power stability, frequency stability, phase noise, qubit coherence and SPAM errors. Microwave $\pi/2$ error estimates were extracted using process matrix analysis to simulate the Ramsey experiment results as described in Section \ref{sec.ErrorProp} and were found to be 0.0028 per atom per microwave $\pi/2$ pulse.  

\subsection{b.2 Local Stark shifted microwave pulses}

To perform a local $x$-axis rotation gate  with a pulse area of $\phi$   we use the decomposition  
$$
R_x(\phi)=R_y(\pi/2)R_z(\phi)R_y(-\pi/2).
$$
The $R_y$ operations are global microwave pulses with a $\pi/2$ phase shift controlled by a signal generator and the  $R_z(\phi)$ is a local Stark shifting pulse which we apply with the 459 nm Rydberg beam. A site which receives no Stark pulse only experiences two microwave rotations that cancel. For $R_x(\pi/2)$ we need   $R_z(\pi/2)$ which was provided by  a $86~\rm  ns$ duration pulse from the 459 nm Rydberg laser. 

The local gate in the Bell state preparation sequence is $R_\theta(\pi/2)$, a $\pi/2$ rotation about an angle $\theta$ in the azimuthal plane and has the form
\begin{align}
\textbf{$R_\theta(\pi/2)$} &= \frac{1}{\sqrt{2}}
\begin{pmatrix}
1 \hfill & i e^{-i \theta} \hfill  \\
 i e^{i \theta} \hfill & 1 \hfill 
\end{pmatrix}.
\end{align}
 The pulse sequence is    
\begin{eqnarray}
\textbf{$R_\theta(\pi/2)$}&=&
\textbf{$R_{\theta-\pi/2}(\pi /2)$}\textbf{$R_{z}(\pi/2)$}\textbf{$R_{\theta+\pi/2}(\pi /2)$}
\end{eqnarray}

This requires phase shifting the microwave pulses which was accomplished with a computer controlled signal generator.
The error for this operation is larger than for the global gates due to noise of the focused laser beam\cite{Xia2015}. Error estimates are calculated from microwave Ramsey experiments with a local $\pi/2$ gate on the target atom.  These estimates are extracted using process matrix analysis to simulate the Ramsey experiment error as described in Sec. \ref{sec.ErrorProp}.  Error estimates for the microwave $\pi/2$ pulses were found to be 0.0028 per atom per pulse (same as above) and the local Stark pulse introduces an additional 0.006 error probability.

\section{SPAM errors}
\label{sec.SPAMErrors}
\subsection{c.1 Atom readout loss}

Atom preparation errors are due to retention loss during measurement to verify an atom is present. Mean and median values across the array are given in Fig. 1 in the main text.  
The measured loss for control and target sites was 0.005 for each site.  Splitting this error between the two readouts yields 0.0025 per atom per readout. 

This loss is partly due to measurement induced loss and partly due  to background collisions at finite vacuum pressure.  Our measured vacuum limited trap lifetime is about 30 s which implies a loss of 0.003 for a 100 ms duration double measurement. This loss contribution can be made negligible with faster measurements and improved vacuum conditions.

\subsection{c.2 Optical pumping}
\label{subsec.OP}
The optical pumping error is difficult to disentangle from the 
atom preparation error since both contribute to a finite amplitude of microwave pulses. One measure of the pumping infidelity is the ratio of pumping to depumping times for preparation of $m_f=0$ clock states. We have observed this ratio as low as $1/200$ corresponding to an error of 0.005 per atom.

\subsection{c.3 State measurement}

State measurement errors were estimated by making Gaussian fits to the observed count distributions for atoms present and not present after pushing out atoms in $f=4$  and defining
the measurement error as the overlap of the distributions. The mean and median errors across the array are given in Fig. 1 in the main text. For the specific sites and measurement times used for the Bell state experiments we obtained lower measurement errors of $5\times 10^{-5}$ for the control site
and $1\times 10^{-4}$ for the target site.  We have used the average value of these errors in the table. 

There is an additional error from blowing away atoms in $f=4$, which may be less than 100\% successful, may depump atoms to $f=3$ instead of removing them from the trap, and may also cause loss of atoms in $f=3$ with very low probability. On the basis of related studies with non-destructive state measurements\cite{Kwon2017} we estimate the errors from blow away are not larger than the overlap error derived from fitting the count distributions. This error estimate was identified to be $1.5 \times 10^{-4}$ per atom.

\section{Process Matrix Error Modeling}
\label{sec.ErrorProp}

To accurately simulate how each of the errors calculated above affects the created Bell state, we use $\chi$-matrices to simulate the quantum process which the atoms experience at each stage of the experiment.  To accomplish this, we break the error model into four segments: 1. state preparation, 2. $C_{Z}$ gate, 3. Local $\pi/2$ rotation on target atom, 4. parity oscillation pulse and state measurement.  There are multiple processes which act on the atoms in each segment, and we can determine how each process affects the atoms through a combination of process modeling and experimental measurement of the various quantum processes.

\subsection{State Preparation}
To prepare atoms in the desired state to be entangled by the $C_{Z}$ gate, there are three distinct steps.  The first of these steps is a readout measurement to determine the occupancy of the array.  Since atoms are loaded into the array probabilistically, it is necessary to post select on situations when both control and target atom sites are loaded.  During the readout process, there is a finite chance that the atom will be lost through either a background collision or by being heated out of the trap.  To determine this probability, we performed a high statistics measurement with two readouts back to back from each other.  During these measurements, there was $0.5\%$ chance that the atom was lost in each of the control and target sites.  Dividing this loss equally between the two readouts, we determine that there is a $0.25\%$ chance of loss per atom per readout.  This loss is modeled as a single qubit loss quantum process matrix.

The next step in state preparation is optically pumping the atoms into the correct state.  As discussed in Sec. \ref{subsec.OP}, the magnitude of this effect is determined from depumping measurements.  To model imperfect optical pumping with quantum process matrices, we must consider how imperfect optical pumping affects the atomic state.  In our optical pumping process, we end each optical pumping cycle by turning off the pumping beam and leaving the repumping on to make sure that all atoms are in the $f=4$ hyperfine ground state.  Any atoms which were not pumped into $\ket{1}=\ket{4,0}$ will end up occupying different magnetic substates in $f=4$.  These states will not be driven by microwave rotations and will be blown away in the final state-selective readout.  Given these effects, we model imperfect optical pumping as loss.  We note that this modeling is a slight approximation since it is possible for a control atom in the wrong magnetic state of $f=4$ to be excited to the Rydberg state and thus blockade the target atom from being excited to the Rydberg state.  Full modeling of this effect would require a higher dimensional quantum process analysis, which is beyond the scope of this model.  As mentioned in Sec.  \ref{subsec.OP}, the loss out of the computational basis due to  imperfect optical pumping was determined to be approximately $0.5\%$ per atom.

Once pumped into $\ket{1}$, both atoms need to be rotated to $(\ket{0}+\ket{1})/\sqrt2$.  This rotation is performed with a microwave pulse resonant with the energy splitting between the two qubit states.  As mentioned in Sec. \ref{subsec.uwave}, several factors can cause errors in microwave rotation.  This error comes in as dephasing about the rotation axis on the Bloch sphere (i.e. states experiencing a rotation will decohere, but states located at the poles of the rotation axis will remain unaffected).  The $\chi$-matrix for an imperfect microwave rotation is given below.  The magnitude of this effect was determined from a microwave Ramsey experiment with a small gap time ($t_{\rm gap}<<T_{2,\rm ground}$).  By varying the phase of the second microwave $\pi/2$ pulse with a constant gap time, a Ramsey curve was measured.  Both readout and optical pumping errors discussed above are present in this Ramsey experiment in addition to the microwave dephasing errors which are being measured.  To extract the microwave dephasing, we model this Ramsey experiment with quantum process matrices.  The microwave dephasing probability can then be extracted from the measured Ramsey curve and was found to be $0.28\%$ per atom per $\pi/2$ microwave pulse.    

\subsection{$C_Z$ Gate}
\label{subsec.CZErrorProp}
The $C_Z$ gate is comprised of three Rydberg pules: 1. A $\pi$ Rydberg pulse on the control atom, 2. A $2\pi$ pulse on the target atom, 3. A final $\pi$ pulse on the control atom.  There are many error sources associated with the $C_Z$ gate and care must be taken to model these errors in the proper order to correctly determine their effect on the atoms.  These errors are listed in Table \ref{tab.budget2} and have been calculated above as well.  Some of these calculated errors have a reasonably high degree of certainty associated with them because they rely on experimental parameters that can be directly measured (e.g. the laser detuning from the $7p_{1/2}$ state can be spectroscopically measured).  However, some of the error calculations have large uncertainties associated with them due to imperfect knowledge of experimental parameters (e.g., atom confinement in the trap cannot be directly measured due to diffraction limitations of the imaging optics).  More accurate values for many of these errors were extracted from various experiments on the atoms.  Also note that some of these error sources depend on the state of both of the atoms and must be modeled by 2-qubit quantum process tensors.  For example, when an atom is excited to the Rydberg state, both lasers contribute to scattering from the $7p_{1/2}$ state; however, if the target atom is blockaded by the control atom, it will not be excited to the Rydberg state, and only the 459 nm laser will contribute to intermediate state scattering.  Matrices and tensors for all quantum processes and how they affect the input quantum state are listed below.

To determine errors arising from laser noise and atom position in Rydberg beams more precisely, we used  $2\pi$ Rabi oscillation  experiments between ground and Rydberg states.  To extract these values, experiments were modeled using parameters defined above in addition to modeling the Rydberg pulses.  We note that optical pumping errors were not included in this modeling with the same magnitude as above since other magnetic sublevels will also experience Rydberg rotations; these states will rotate at different rates than $\ket{1}$\cite{Maller2015}, so they do not entirely return to the ground state in the $2\pi$ Rydberg pulse.  Also, Rydberg state detection is not perfect, so the observed $2\pi$ retention is slightly higher than it should be.  From modeling these experiments, we estimate that there is approximately a $0.5\%$ error per Rydberg $\pi$ pulse due to the combined effects from laser noise and atom position.  

Since laser noise and atom position variations have a similar effect on the atomic state (namely leaving the atom in the Rydberg state), they cannot be distinguished without further measurements.  Servo bumps were measured from a heterodyne experiment as discussed in Sec. \ref{subsec.LaserNoise} and their effect was numerically estimated to be $0.5\%$ in a $2\pi$ pulse.  The remaining $0.5\%$ we attribute to atom position variations in the Rydberg beams.  We note that this value is significantly lower than estimated in Sec. \ref{subsec.Spatial} which  points to a tighter atom localization in the trap than estimated from trap parameters.  The exact origin of this discrepancy is unknown. However, we hypothesize that it is due to spatial filtering and/or aberrations of the blue-detuned trap broadening out the trap lines.  This results in tighter confinement.  We have seen some evidence of this effect in higher than expected trap frequencies in parametric heating experiments and will investigate this further in future studies.  

Both laser noise and atom position variation leave some atom population in the Rydberg state which will be primarily lost.  Therefore, we can model these sources of error as state dependent loss, i.e., if an atom is in the $\ket{1}$ state, some population might be lost, but if the atom is in the $\ket{0}$ state, the Rydberg lasers are non-resonant, and the atom will not be affected by this error.  Similarly, if the control atom is excited to the Rydberg state, the target atom will be blockaded from being excited to the Rydberg state and will not experience these errors.  Therefore, we model both laser noise and atom position variation under the Rydberg beams as a state dependent loss process for the control atom and as a controlled state dependent loss process for the target atom.   

Rydberg crosstalk (excitation of neighboring sites) can be determined from Rydberg Rabi oscillation data for the control site (by post selecting cases when the control atom was not loaded into the lattice). From these experiments, we see that there was less than or equal to $0.25\%$ excitation per neighboring atom.  During the $C_Z$ gate, this unintended Rydberg excitation mainly affects the Rydberg excitation of the target atom.  Specifically, when the control atom is in $\ket{0}$, the Rydberg pulse on the control atom does not excite the control, but could possibly excite neighboring atoms in state $\ket{1}$  to the Rydberg state.  This unintended excitation would blockade the target atom leading to an incorrect phase.  There is a small probability that neighboring atoms would be excited when the control atom is in the $\ket{1}$ state; however, this probability is very small compared to the first effect discussed and has little influence on the gate performance.  Since the effect of crosstalk on the gate depends on the state of both atoms, it must be modeled as a two-qubit quantum process.

The finite lifetime of the Rydberg state only significantly affects the control atom.  The calculation performed in Sec. \ref{subsec.Lifetime} relies on well-known experimental parameters. However, since the control atom blockades the target atom in the $C_{Z}$ gate, a distinction must be made for decay before the target Rydberg pulse or after.  Since the target pulse is centered in time between the two control Rydberg pulses, we model half of the decay from the Rydberg state before the Rydberg pulse on the target and half of the decay after.  This decay is modeled as loss because decay will be primarily back into the $f=4$ state, which will then be excited to the Rydberg state by the second control pulse.  Population left in the Rydberg state will be primarily lost due to ionization from subsequent microwave pulses in the Bell state preparation sequence.  This decay only occurs when the control atom has probability amplitude in the $\ket{1}$ state going into the $C_Z$ gate, so it is included in calculations as a state-dependent loss.

Doppler dephasing also acts primarily on the control atom, dephasing the superposition of $\ket{0}$ and the Rydberg state.  Since this process does not alter the control atom's probability of Rydberg excitation, the target atom will not be influenced by this effect (the target spends a much smaller amount of time in the Rydberg state). Since the Rydberg population returns to $\ket{1}$, this Doppler dephasing acts as dephasing between $\ket{0}$ and $\ket{1}$ qubit states.  As seen in Sec. \ref{subsec.Doppler}, the Doppler dephasing error magnitude is determined by the atom temperature, which can be extracted from trap-drop and recapture experiments as well as qubit ground Ramsey experiments.  Both of these experiments are used to deduce an atom temperature of $15 ~\mu \rm K$, which indicates a dephasing probability of $1.3\%$.

After accounting for all the error sources listed above, we simulated the eye diagram results (see main text Fig. 4).  In this experiment, the target is placed in an equal superposition of the two ground states, specifically $(\ket{0}+\ket{1})/\sqrt2$, using a local gate, and then the $C_Z$ gate sequence is performed on Rydberg and target sites.  When the control atom is loaded, the atom is blockaded and does not get excited to the Rydberg state. However, when the control atom is not loaded, the atom gets excited to the Rydberg state and acquires a $\pi$ phase shift with respect to the previous case.  In our measurements, we noticed that the non-blockaded case experiences a significant amount of additional dephasing compared to the blockaded case.  Our models do not explain this additional dephasing term; however, we need to include this effect to make our model consistent.  We have extracted the value of this dephasing term by simulating the eye diagram experiment using quantum process matrices and determined the dephasing probability that is introduced by the $2\pi$ Rydberg pulse.  If the excitation is blockaded we still observe some dephasing, but only approximately $1/3$ of what we see in the nonblockaded case.  In similar experiments, we have determined that the control atom sees this Rydberg laser dephasing term as well.  We incorporate this into our model as dephasing between $\ket{0}$ and $\ket{1}$ when the atom sees a Rydberg pulse.  Through our simulations, we find this error to be $1.8\%$ per Rydberg $\pi$ pulse and $0.6\%$ if the pulse is blockaded.  During the $2\pi$ excitation on the target, the amount of dephasing depends on whether or not the excitation is blockaded, so two-qubit process modeling is required to properly account for this dephasing term on the target.

\subsection{Local state rotation}
After the state of the atoms is prepared and acted on by the $C_Z$ gate, the two-atom state is entangled. However, the target atom must be rotated by $\pi/2$ before the atoms occupy a Bell state in the computational basis.  To perform this local rotation, we use a Stark shift pulse from the 459 nm laser sandwiched between two microwave rotations with opposite phases.  To include these errors in the experimental model, we model microwave pulses identically to how they are used above. We include an additional error arising from the Stark pulse as dephasing between the two hyperfine ground states.  The value of this error can be extracted from Ramsey experiments, where the first $\pi/2$ rotation is provided by a local gate as described above and the second $\pi/2$ rotation is provided by a global microwave rotation.  By scanning the phase of the second $\pi/2$ rotation, a Ramsey curve is measured, and the amplitude can be extracted.  By modeling this Ramsey experiment with process matrices, we find that the local Stark pulse gives a dephasing error contribution of $0.6\%$.  This local Stark shift only affects the state of the target atom. However, errors from microwave rotations will affect the control atom as well and are included in the error model of the Bell state creation experiment.

\subsection{Bell state measurement}
After the Bell state is created, measurements are made to determine the fidelity.  These consist of two separate measurements, a state population measurement and a parity oscillation.  Both of these measurements consist of processes already discussed above.  The population measurement consists of a single state selective readout.  The parity oscillation measurements consist of a global $\pi/2$ microwave rotation followed by a state-selective readout.  By varying the phase of the microwave rotation, the parity curve is measured.  Both of these two measurement types are included in our error model of the experiment.

\begin{figure*}
    \centering
    \includegraphics[width=1.5\columnwidth]{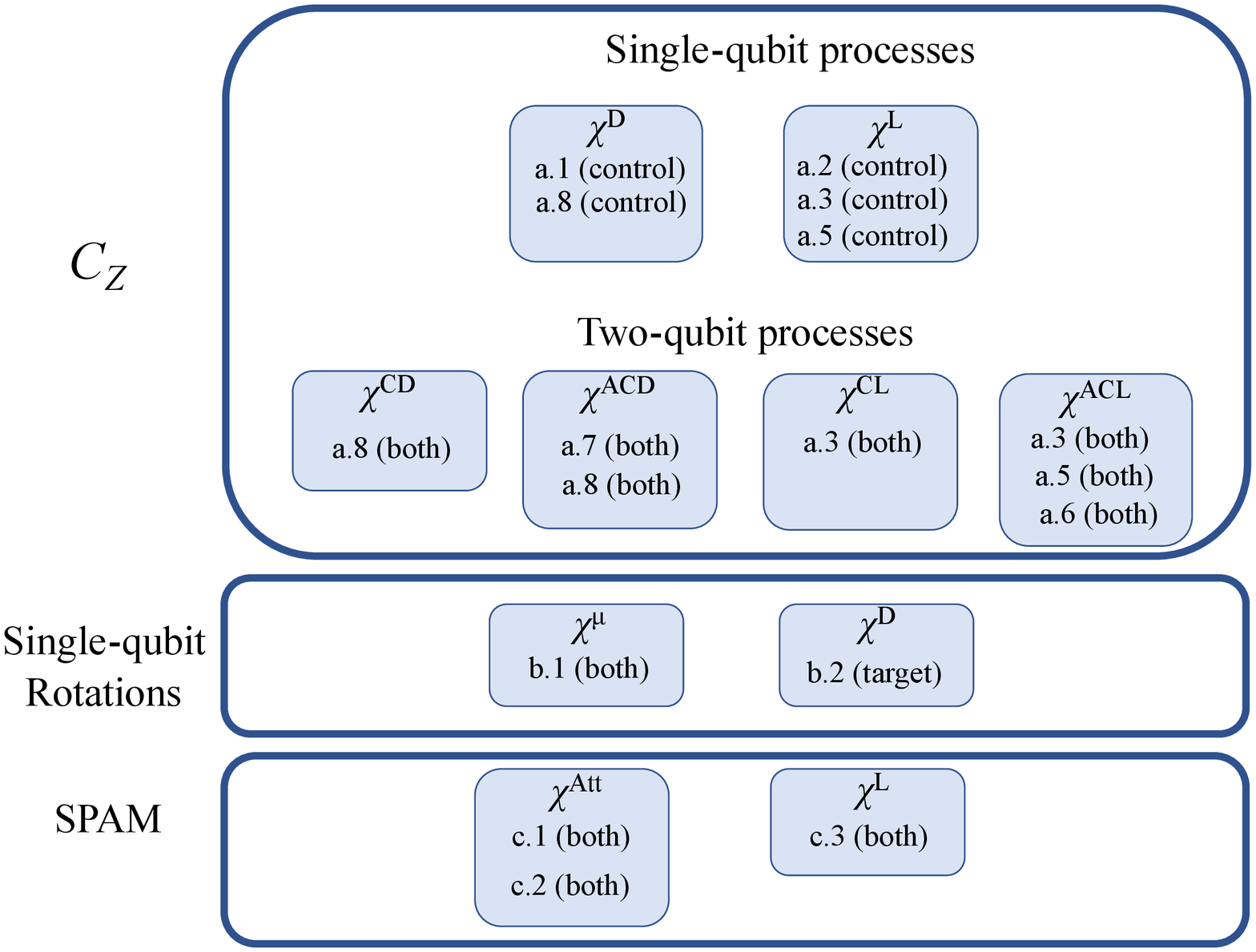}
    \caption{A summary of error sources and how they are incorporated into the quantum process model of the experiment.  The errors are labeled according to the entries in Table \ref{tab.budget2}.  Some errors are listed in more than one process matrix category in the $C_Z$ section; this is due to the fact that the error introduced when the target is excited to the Rydberg state depends on whether or not the control atom is in the Rydberg state.}
    \label{fig.SM Q Processes}
\end{figure*}

\subsection{Error Propagation with Process matrices}
\label{sec.processmatrix}

To properly incorporate all the effects listed above into an error model of the experiment, we use one- and two-qubit $\chi$-matrix propagation of errors.  Many of the processes used for Bell state preparation act on the atomic states separately.  Because these processes act separably on the two atoms, they can be modeled with single qubit $\chi$-matrices.  All of the 
$\chi$ matrices are listed in Fig.  \ref{fig.SM Q Processes}.

Processes acting on the control atom take the following form:

\begin{equation}
\label{eqn.SingleQubitProcess}
    {\mathcal E}^j(\rho) = \sum_{m,n=1}^{4}\chi_{m,n}^j(\sigma_m\otimes I) \rho (\sigma_n\otimes I),
\end{equation}

Where $\rho$ is the input quantum state, ${\mathcal E^j}(\rho)$ is the output state after it has been acted on by quantum process $j$, $\sigma\in\{I,\sigma_x,\sigma_y,\sigma_z\}$ where $\sigma_{x,y,z}$ represent $x$, $y$ or $z$, Pauli matrices respectively, and $\chi$ is a quantum process matrix which contains all the quantum process information and weights each combination of Pauli matrices in the sum.  Similarly, quantum processes acting on the target atom take the following form
\begin{equation}
    {\mathcal E}^{j}(\rho) = \sum_{m,n=1}^{4} \chi_{m,n}^{j} (I\otimes \sigma_m) \rho (I\otimes \sigma_n ),
\end{equation}
Several single qubit quantum processes are required to simulate the Bell state creation and measurement demonstration as listed in the previous subsections. 

In the following we  list $\chi$ matrices for the one-qubit processes used in our analysis.  The single-qubit $\chi$ matrix for the identity process is
\begin{align}
\textbf{$\chi^{\rm I}$} &=
\begin{pmatrix}
1 \hfill & 0 \hfill & 0 \hfill & 0 \hfill \\
0 \hfill & 0 \hfill & 0 \hfill & 0 \hfill \\
0 \hfill & 0 \hfill & 0 \hfill & 0 \hfill \\
0 \hfill & 0 \hfill & 0 \hfill & 0 \hfill
\end{pmatrix}.
\end{align}
The single qubit-process matrix for partial loss/attenuation (probability $\epsilon$) of the quantum state is
\begin{align}
\textbf{$\chi^{\rm Att}$} &=
\begin{pmatrix}
1-\epsilon \hfill & 0 \hfill & 0 \hfill & 0 \hfill \\
0 \hfill & 0 \hfill & 0 \hfill & 0 \hfill \\
0 \hfill & 0 \hfill & 0 \hfill & 0 \hfill \\
0 \hfill & 0 \hfill & 0 \hfill & 0 \hfill
\end{pmatrix}.
\end{align}
The single qubit process matrix for partial dephasing (probability $\epsilon$) of the quantum state is
\begin{align}
\textbf{$\chi^{\rm D}$} &=
\begin{pmatrix}
1-\epsilon \hfill & 0 \hfill & 0 \hfill & 0 \hfill \\
0 \hfill & 0 \hfill & 0 \hfill & 0 \hfill \\
0 \hfill & 0 \hfill & 0 \hfill & 0 \hfill \\
0 \hfill & 0 \hfill & 0 \hfill & \epsilon \hfill
\end{pmatrix}.
\end{align}
The single qubit-process matrix for state-dependent loss (probability $\epsilon$) of the quantum state is
\begin{align}
\textbf{$\chi^{\rm L}$} &=
\begin{pmatrix}
1-3\epsilon/4 \hfill & 0 \hfill & 0 \hfill & \epsilon/4 \hfill \\
0 \hfill & 0 \hfill & 0 \hfill & 0 \hfill \\
0 \hfill & 0 \hfill & 0 \hfill & 0 \hfill \\
\epsilon/4 \hfill & 0 \hfill & 0 \hfill & \epsilon/4 \hfill
\end{pmatrix}.
\end{align}
Finally the single-qubit process matrix for a $\pi/2$ microwave rotation about the $x$-axis of the Bloch sphere with dephasing (probability $\epsilon$) during the rotation is
\begin{align}
\textbf{$\chi^{ \mu}$} &=
\begin{pmatrix}
1/2 \hfill & 0 \hfill & i(1/2-\epsilon) \hfill & 0 \\
0 \hfill & 0 \hfill & 0 \hfill & 0 \hfill \\
i(-1/2+\epsilon) \hfill & 0 \hfill & 1/2 \hfill & 0 \hfill \\
0 \hfill & 0 \hfill & 0 \hfill & 0 \hfill
\end{pmatrix}.
\end{align}
For microwave rotations about an axis on the equator that is at an angle $\theta$ with respect to the $x$-axis, a unitary transformation is used to rotate the input state by an amount $\theta$ before the quantum process and to rotate by $-\theta$ after the quantum process.  This unitary operator has the form

\begin{align}
\textbf{$U_z(\theta)$} &= 
\begin{pmatrix}
1 \hfill & 0 \hfill  \\
0 \hfill & e^{i \theta} \hfill 
\end{pmatrix}.
\end{align}

In addition to the single qubit quantum processes listed above, two-qubit quantum processes are required to simulate how some errors affect an input quantum state during the $C_Z$ gate.  A two-qubit quantum process acts on an input state $\rho$ using the following rule:

\begin{equation}
    {\mathcal E}^j(\rho) = \sum_{m,r,n,s=1}^{4} \chi_{m,r,n,s}^j (\sigma_m \otimes \sigma_r) \rho (\sigma_n \otimes \sigma_s),
\end{equation}

where variable definitions are the same as in Eq. (\ref{eqn.SingleQubitProcess}) except now $\chi_{m,r,n,s}$ is a four dimensional tensor that contains all the information of the two-qubit quantum process and weights all combinations of Pauli matrices in the summation.  Below we  list the $\chi$ matrices for the two-qubit processes used in our analysis.  

The two-qubit process matrix for controlled partial dephasing (i.e. the target experiences partial dephasing if the control is in $\ket{1}$) is
\begin{eqnarray}
\textbf{$\chi^{\rm CD}$} &&= 
(1-\epsilon)\chi^{I}\otimes\chi^{I}\nonumber\\
&&+
\epsilon
\begin{pmatrix}
1/2 \hfill & 0 \hfill & 0 \hfill & 1/2 \hfill \\
0 \hfill & 0 \hfill & 0 \hfill & 0 \hfill \\
0 \hfill & 0 \hfill & 0 \hfill & 0 \hfill \\
1/2 \hfill & 0 \hfill & 0 \hfill & -1/2 \hfill 
\end{pmatrix}
\otimes
\begin{pmatrix}
1/2 \hfill & 0 \hfill & 0 \hfill & 1/2 \hfill \\
0 \hfill & 0 \hfill & 0 \hfill & 0 \hfill \\
0 \hfill & 0 \hfill & 0 \hfill & 0 \hfill \\
1/2 \hfill & 0 \hfill & 0 \hfill & -1/2 \hfill 
\end{pmatrix}.
\end{eqnarray}
The two-qubit process matrix for anti-controlled partial dephasing (i.e. the target experiences partial dephasing if the control is in $\ket{0}$) is
\begin{eqnarray}
\textbf{$\chi^{\rm ACD}$} &&= 
(1-\epsilon)\chi^{I}\otimes\chi^{I}\nonumber\\
&&+
\epsilon
\begin{pmatrix}
1/2 \hfill & 0 \hfill & 0 \hfill & 1/2 \hfill \\
0 \hfill & 0 \hfill & 0 \hfill & 0 \hfill \\
0 \hfill & 0 \hfill & 0 \hfill & 0 \hfill \\
-1/2 \hfill & 0 \hfill & 0 \hfill & 1/2 \hfill 
\end{pmatrix}
\otimes
\begin{pmatrix}
1/2 \hfill & 0 \hfill & 0 \hfill & 1/2 \hfill \\
0 \hfill & 0 \hfill & 0 \hfill & 0 \hfill \\
0 \hfill & 0 \hfill & 0 \hfill & 0 \hfill \\
-1/2 \hfill & 0 \hfill & 0 \hfill & 1/2 \hfill 
\end{pmatrix}.
\end{eqnarray}
The two-qubit process matrix for controlled partial polarization (i.e. the target experiences partial polarization if the control is in $\ket{1}$) is
\begin{eqnarray}
\textbf{$\chi^{\rm CL}$} &&= 
(1-\epsilon)\chi^{I}\otimes\chi^{I}\nonumber\\
&&+
\epsilon
\begin{pmatrix}
3/4 \hfill & 0 \hfill & 0 \hfill & 1/4 \hfill \\
0 \hfill & 0 \hfill & 0 \hfill & 0 \hfill \\
0 \hfill & 0 \hfill & 0 \hfill & 0 \hfill \\
1/4 \hfill & 0 \hfill & 0 \hfill & -1/4 \hfill 
\end{pmatrix}
\otimes
\begin{pmatrix}
3/4 \hfill & 0 \hfill & 0 \hfill & 1/4 \hfill \\
0 \hfill & 0 \hfill & 0 \hfill & 0 \hfill \\
0 \hfill & 0 \hfill & 0 \hfill & 0 \hfill \\
1/4 \hfill & 0 \hfill & 0 \hfill & -1/4 \hfill 
\end{pmatrix}.
\end{eqnarray}
The two-qubit process matrix for anti-controlled partial polarization (i.e. the target experiences partial polarization if the control is in $\ket{0}$) is
\begin{eqnarray}
\textbf{$\chi^{\rm ACL}$} &&= 
(1-\epsilon)\chi^{I}\otimes\chi^{I}\nonumber\\
&&+
\epsilon
\begin{pmatrix}
3/4 \hfill & 0 \hfill & 0 \hfill & 1/4 \hfill \\
0 \hfill & 0 \hfill & 0 \hfill & 0 \hfill \\
0 \hfill & 0 \hfill & 0 \hfill & 0 \hfill \\
-1/4 \hfill & 0 \hfill & 0 \hfill & 1/4 \hfill 
\end{pmatrix}
\otimes
\begin{pmatrix}
3/4 \hfill & 0 \hfill & 0 \hfill & 1/4 \hfill \\
0 \hfill & 0 \hfill & 0 \hfill & 0 \hfill \\
0 \hfill & 0 \hfill & 0 \hfill & 0 \hfill \\
-1/4 \hfill & 0 \hfill & 0 \hfill & 1/4 \hfill 
\end{pmatrix}.
\end{eqnarray}

\begin{figure*}[!t]
\vspace{-.0cm}
\centering
 \includegraphics[width=2\columnwidth]{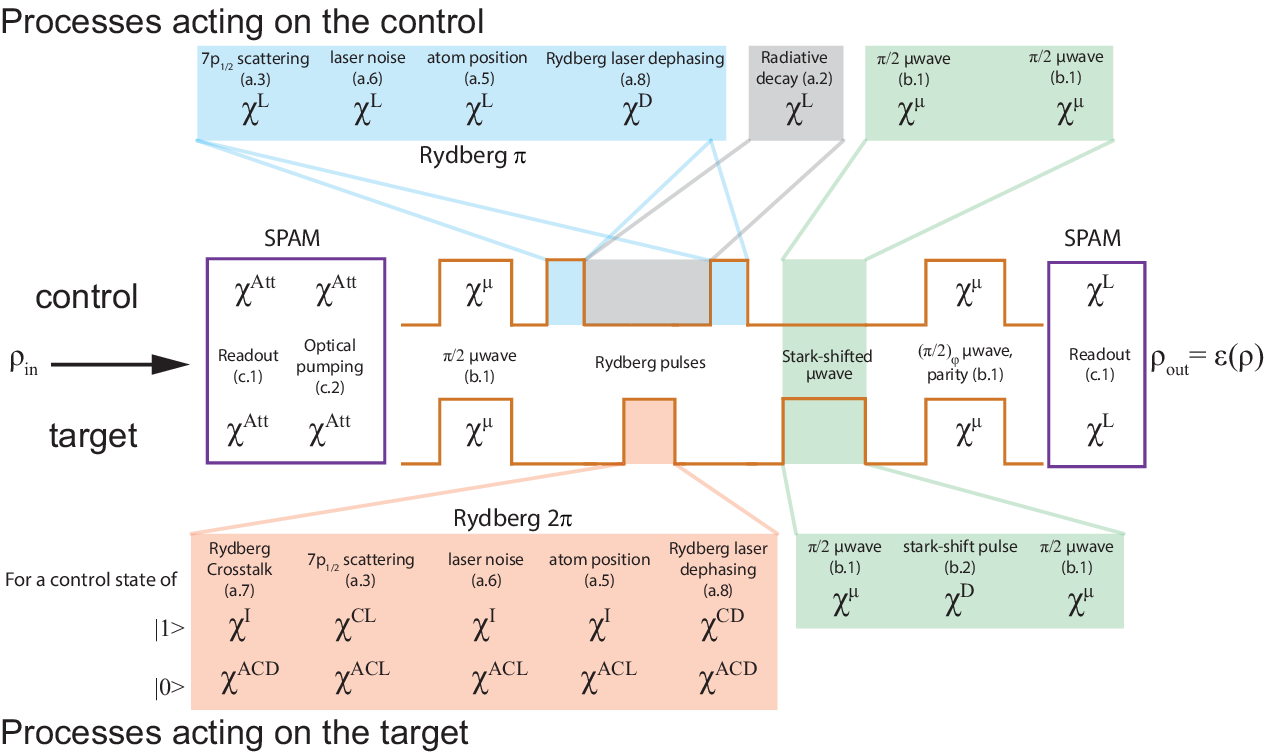}
  \caption{Visual representation of the error channels and corresponding process matrices.}
\label{figsm4_QPPropagation}
\end{figure*}

By propagating the input quantum state using process matrices, we are able to properly weight the contributions of each of the errors using the steps shown in Fig. \ref{figsm4_QPPropagation}.  After quantum process propagation, the fidelity, $F_{\rm Bell}$, of the output state is calculated relative to  an ideal Bell state output
\begin{equation}
    F_{\rm Bell} = \left( \rm Tr \left[ \sqrt{\sqrt{\rho_{\rm Bell}} \rho_{\rm out}\sqrt{\rho_{\rm Bell}}} \right] \right) ^2 ,
\end{equation}
where $\rho_{Bell}$ and and $\rho_{out}$ represent the density matrices for an ideal Bell state and the modeled output state, respectively.  Given the input errors, our model predicts $F_{\rm Bell}=85.3\%$, which agrees well with our measured result. Setting  errors c.1-c.3 to zero, we calculate $F_{\rm Bell}^{\rm -SPAM} = 87.7\%$. Finally, also setting errors b.1 and b.2 to zero, we calculate the fidelity of the output Bell state accounting only for the  errors  introduced by the $C_Z$ gate, $F_{\rm Bell}^{C_Z}=88.7\%$.  

We note that this model for error propagation, while very detailed,  has a few limitations.  In particular, error propagation using $\chi$-matrices does not allow for non-Markovian processes, i.e. processes where an input state increases in coherence.  Such processes occur in spin-echo sequences and could be present in this demonstration during local microwave rotations and/or during $\pi$-gap-$\pi$ processes on the control atom.  We have assumed that such effects are small compared to the total error model.  Full treatment of this process could be performed by master equation simulation of the full Bell state creation experiment; however, such analysis is beyond the scope of this work.

\begin{widetext}

\section{Decoherence of Rabi oscillations}
\label{sec.rabiramsey}

The $T_2$ coherence time of a qubit measured in a Ramsey interference experiment is generally much shorter
than the observed decoherence time of a driven qubit performing Rabi oscillations.  
We calculate the decoherence of a driven qubit starting  with the initial mixed state 
$$
\rho(0)=\ket{0}\bra{0}\otimes \sum_\bfn P_\bfn \ket{\bfn}\bra{\bfn}
.$$
Here $\bfn=(n_x,n_y,n_z)$ denotes a vibrational state with excitations $n_x,n_y,n_z$
along the Cartesian axes. 
The state after Rabi oscillations for a time $t$ is 
$$
\rho(t)=U \rho(0)U^\dag
$$ 
with 
$$
U=\sum_{\bfn}R(t,\bfn)\ket{\bfn}\bra{\bfn}.
$$
Here we allow the rotation operator to depend on $\bfn$. 

The transformed state is 
\bea
\rho(t)&=& \sum_{\bfn'}R(t,\bfn')\ket{\bfn'}\bra{\bfn'}
\left(\ket{0}\bra{0}\otimes \sum_\bfn P_\bfn \ket{\bfn}\bra{\bfn} \right)
\sum_{\bfn''}R^\dag (t,\bfn'')\ket{\bfn''}\bra{\bfn''}\nn\\
&=&
\sum_{\bfn}R(t,\bfn)\ket{0}\bra{0} R^\dag (t,\bfn)
P_\bfn \ket{\bfn}\bra{\bfn}\nn
\eea
and the probability of observing $\ket{1}$ after time $t$ is 
\bea
P_{\ket{1}}(t)&=&{\rm Tr}_{\rm vib}\left[\bra{1}\rho(t) \ket{1} \right]\nn\\
&=&\sum_{\bfn'} \bra{\bfn'}\bra{1}\rho(t) \ket{1}\ket{\bfn'}\nn\\
&=&\sum_{\bfn}P_\bfn| \bra{1} R(t,\bfn)\ket{0}|^2.\nn
\eea
The matrix element is 
\bea
| \bra{1} R(t,\bfn)\ket{0}|^2&=&\frac{|\Omega|^2}{|\Omega|^2+\Delta_1^2(\bfn)}\sin^2\left(\frac{\left(|\Omega|^2+\Delta_1^2(\bfn)\right)^{1/2}t}{2} \right),\nn\\
&=&\frac{|\Omega|^2}{2\left[|\Omega|^2+\Delta_1^2(\bfn)\right]}\left[1-\cos\left(\left(|\Omega|^2+\Delta_1^2(\bfn)\right)^{1/2}t\right)\right]\nn
\eea
with $\Omega$ the Rabi frequency and $\Delta_1(\bfn)$ the vibrational state dependent detuning of the qubit frequency. For an atomic qubit in an optical trap this shift is due to  
differential light shifts that depend on the wavelength of the trap light\cite{Rosenbusch2009,Carr2016}.

We can extract the scaling of the coherence time with temperature from an approximate analysis. Different vibrational states will dephase when 
$$
\left[\left(|\Omega|^2+\Delta_1^2(\bfn)\right)^{1/2}-|\Omega|\right]t\sim\pi
$$
so
$$
T_2\sim \frac{2\pi|\Omega|}{\Delta_1^2(\bfn)}.
$$
Since $\Delta_1(\bfn)\sim \bfn$ and $\langle \bfn \rangle\sim T$ in the thermal limit we see that $T_2\sim |\Omega|/T^2$.

\subsection{Semiclassical approximation}

We can extract useful expressions from a semiclassical approximation.
Using 
$$
P_\bfn\simeq \omega_x\omega_y\omega_z \beta^3 e^{-\beta \bfn\cdot\bfomega}
$$
with $\beta=\hbar/k_B T$ 
 we find 
\bea
P_{\ket{1}}(t)
&=&-\frac{\beta^3 \omega_x\omega_y\omega_z}{2}\iiint_0^\infty dn_x dn_ydn_z\, e^{-\beta \bfn\cdot\bfomega }
\frac{|\Omega|^2}{|\Omega|^2+(\frac{\bar\Delta_{\rm LS}}{2}\bfn\cdot\bfomega)^2}\nn\\
&&\times \cos\left[\left(|\Omega|^2+(\frac{\bar\Delta_{\rm LS}}{2}\bfn\cdot\bfomega)^2\right)^{1/2} t\right]\nn.
\eea
Here we have used $\Delta_1(\bfn)=\frac{\bar\Delta_{\rm LS}}{2}\bfn\cdot\bfomega$ with $\bar\Delta_{\rm LS}$ the fractional differential light shift of the transition. 
We then make the approximation that $|\Omega|^2\gg (\frac{\bar\Delta_{\rm LS}}{2}\bfn_{\rm max}\cdot\bfomega)^2$ to write the right hand side as
\bea
a&=&\frac{\bar\Delta_{\rm LS}^2\beta^3 \omega_x\omega_y\omega_z}{8|\Omega|^2}\iiint_0^\infty dn_x dn_ydn_z\, e^{-\beta \bfn\cdot\bfomega }
(\bfn\cdot\bfomega)^2
 \cos\left[|\Omega| t + \frac{\bar\Delta_{\rm LS}^2}{8|\Omega|}
(\bfn\cdot\bfomega)^2 t\right]\nn\\
&=&\frac{\bar\Delta_{\rm LS}^2\beta^3 \omega_x\omega_y\omega_z}{8|\Omega|^2}\iiint_0^\infty dn_x dn_ydn_z\, e^{-\beta \bfn\cdot\bfomega }
(\bfn\cdot\bfomega)^2\nn\\
&&\times \left\{ 
 \cos(|\Omega| t)\cos\left[\frac{\bar\Delta_{\rm LS}^2}{8|\Omega|}
(\bfn\cdot\bfomega)^2 t\right]
- \sin(|\Omega| t)\sin\left[\frac{\bar\Delta_{\rm LS}^2}{8|\Omega|}
(\bfn\cdot\bfomega)^2 t\right]
\right\}\nn.
\eea

\begin{figure}[!t]
\vspace{-.0cm}
\centering
 \includegraphics[width=11.cm]{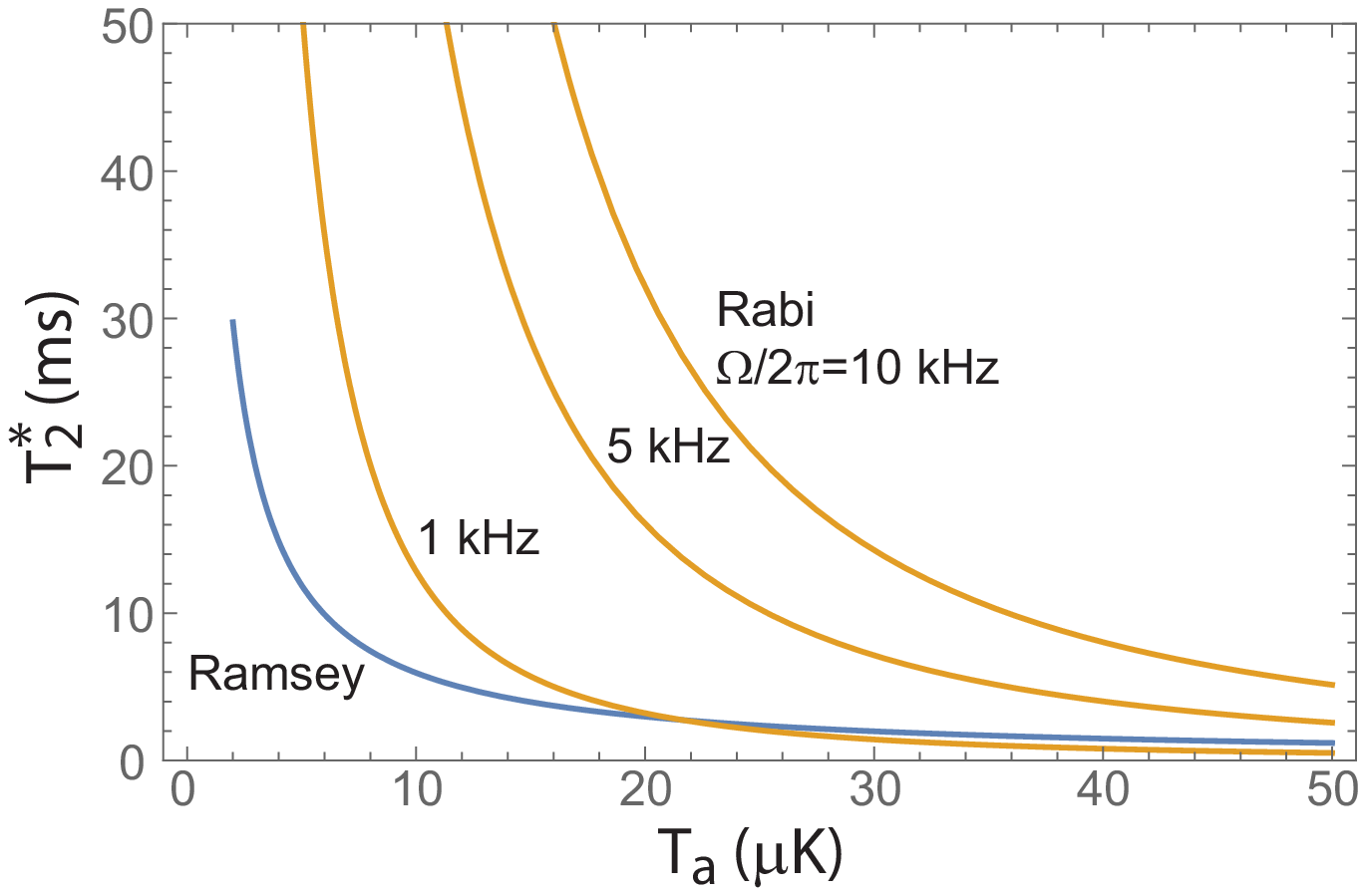}
  \caption{Semiclassical results for $T_2^*$ as measured by Ramsey and Rabi experiments. Parameters $\bar\Delta_{\rm LS}=2.5\times 10^{-4}$.}
\label{fig.T2RandR}
\end{figure}

At $t=0$ we find
\bea
a_0&=&\frac{\bar\Delta_{\rm LS}^2\beta^3 \omega_x\omega_y\omega_z}{8|\Omega|^2}\iiint_0^\infty dn_x dn_ydn_z\, e^{-\beta \bfn\cdot\bfomega }
(\bfn\cdot\bfomega)^2\nn\\
&=&\frac{3\bar\Delta_{\rm LS}^2 }{2|\Omega|^2\beta^2}
\eea
At $t=2mt_\pi=2m\pi/|\Omega|$ we find
\bea
a_m&=&\frac{\bar\Delta_{\rm LS}^2\beta^3 \omega_x\omega_y\omega_z}{8|\Omega|^2}\iiint_0^\infty dn_x dn_ydn_z\, e^{-\beta \bfn\cdot\bfomega }
(\bfn\cdot\bfomega)^2 \cos\left[ \frac{m\pi \bar\Delta_{\rm LS}^2}{4|\Omega|^2}
(\bfn\cdot\bfomega)^2 \right]\nn.
\eea
In addition 
\bea
a_m'&=&\frac{\bar\Delta_{\rm LS}^2\beta^3 \omega_x\omega_y\omega_z}{8|\Omega|^2}\iiint_0^\infty dn_x dn_ydn_z\, e^{-\beta \bfn\cdot\bfomega }
(\bfn\cdot\bfomega)^2 \sin\left[ \frac{m\pi \bar\Delta_{\rm LS}^2}{4|\Omega|^2}
(\bfn\cdot\bfomega)^2 \right]\nn.
\eea

We can define the $T_2^*$ time as the time when $a_m/a_0=1/e$. 
Approximating the $\cos$ term by a 2nd order expansion we find
$$
a_m = a_0-\frac{315 \pi^2 m^2 }{4}\left(\frac{\bar\Delta_{\rm LS}}{|\Omega|\beta}\right)^6=a_0-\frac{315  }{16}\left(\frac{\bar\Delta_{\rm LS}}{|\Omega|^{2/3}\beta}\right)^6t^2.
$$
and to first order in $t$
$$
a_m'=\frac{45   }{8}\left(\frac{\bar\Delta_{\rm LS}}{|\Omega|^{3/4}\beta}\right)^4 t .
$$

 The coherence time is then 
\be
T_{2,\rm Rabi}^*=\left(\frac{8(1-1/e)}{105} \right)^{1/2}\frac{|\Omega|}{\bar\Delta_{\rm LS}^2}\beta^{2}= 0.219 \frac{\hbar^2|\Omega| }{\bar\Delta_{\rm LS}^2 k_B^2 T^2}.
\label{eq.T2Rabi}
\ee
We see that the driving frequency extends the coherence time proportional to $|\Omega|$ but the temperature dependence is $1/T^2$ which is steeper than the $1/T$ scaling of the Ramsey coherence in an optical trap\cite{Kuhr2005}. 
The coherence time for Ramsey and Rabi experiments using parameters relevant for microwave driving in an optical trap are compared in Fig. \ref{fig.T2RandR}. An analogous result holds for the effective observed coherence of the ground-Rydberg Rabi oscillations
shown in Fig. 2 in the main text.

\end{widetext}


\begin{thebibliography}{43}
\bibitem [{not()}]{note2019}%
  \BibitemOpen
  \href@noop {} {}\bibinfo {note} {Our expression for $T_{2,\rm D}$ is $\sqrt
  2$ larger than that in \cite{Wilk2010} where this dephasing mechanism was
  first pointed out.}\BibitemShut {Stop}%
\end{thebibliography}
\end{document}